\documentclass{nature}
\usepackage{bm}
\usepackage{amsmath,bbm}
\usepackage{amsmath, amssymb, soul}
\usepackage{graphicx}
\usepackage[linktocpage=true]{hyperref}
\usepackage{pgfplots}

\newcounter{defcounter} 
\newenvironment{myequation}{%
\addtocounter{equation}{-1}
\refstepcounter{defcounter}

\begin{equation}}
{\end{equation}}

\title{Experimental Phase Diagram of a One-Dimensional Topological Superconductor}

\author{Jun Chen$^{1,}$\footnotemark~, Peng Yu$^{1,}$\footnotemark[\value{footnote}]~,
\footnotetext{These authors contributed equally to this work.}
John Stenger$^2$, Mo{\"i}ra Hocevar$^3$, Diana Car$^4$, S{\'e}bastien R. Plissard$^5$,
Erik P.A.M. Bakkers$^{4,6}$, Tudor D. Stanescu$^{2,}$
\footnote{Current address: Condensed Matter Theory Center and Joint Quantum Institute, Department of Physics, University of Maryland, College Park, Maryland 20742-4111, USA}~
\& Sergey M. Frolov$^{1,}$
\footnote{E-mail: frolovsm@pitt.edu}}

\begin{document}

\maketitle

\begin{affiliations}
\item Department of Physics and Astronomy, University of Pittsburgh, Pittsburgh, PA 15260, USA
\item Department of Physics and Astronomy, West Virginia University, Morgantown, WV 26506, USA
\item Institut N{\'e}el CNRS, 38042 Grenoble, France
\item Eindhoven University of Technology, 5600 MB, Eindhoven, The Netherlands
\item LAAS CNRS, Universit{\'e} de Toulouse, 31031 Toulouse, France
\item QuTech and Kavli Institute of Nanoscience, Delft University of Technology, 2628 CJ Delft, The Netherlands

\end{affiliations}

\begin{abstract}
Topological superconductors can host Majorana quasiparticles which supersede the fermion/boson dichotomy and offer a pathway to fault tolerant quantum computation \cite{ReadPRB2000,KitaevAP03,NayakRMP2008}. In one-dimensional systems zero-energy Majorana states are bound to the ends of the topologically superconducting regions\cite{KitaevPU2001}. An experimental signature of a Majorana bound state is a conductance peak at zero source-drain voltage bias in a tunneling experiment\cite{MourikScience2012,Nadj-PergeScience2014}. Here, we identify the bulk topological phase in a semiconductor nanowire coupled to a conventional superconductor\cite{LutchynPRL2010,OregPRL2010}. We map out its phase diagram through the dependence of zero-bias peak on the chemical potential and magnetic field. Our findings are consistent with  calculations for a finite-length topological nanowire\cite{PotterPRB2011,StanescuPRB2011,MishmashPRB2016}. Knowledge of the phase diagram makes it possible to predictably tune nanowire segments in and out of the topological phase, thus controlling the positions and couplings of multiple Majorana bound states. This ability is a prerequisite for Majorana braiding, an experiment in which Majorana quantum states are exchanged in order to both demonstrate their non-abelian character and realize topological quantum bits\cite{AliceaNatphy2011,vanHeckNJP2012}.
\end{abstract}

We use a prescription for generating Majorana bound states (MBS) that includes four ingredients: a one-dimensional quantum wire, with spin-orbit interaction and induced superconductivity, under external magnetic field $B$\cite{LutchynPRL2010,OregPRL2010}. This combination of ingredients induces a topological superconductor when the following condition is satisfied (Fig. 1a):

\begin{equation}
E_{Z} > \sqrt{\Delta^2 + \mu^2}
\label{eq:phase condition}
\end{equation}
where $E_{Z}= g\mu_{B}B$ is the Zeeman energy, with $g$ the effective Land{\'e} $g$-factor, $\mu_{B}$ the Bohr magneton. $\Delta$ is the induced superconducting gap at $B=0$, and $\mu$ is the chemical potential in the quantum wire, with $\mu=0$ set to coincide with the lowest energy of a one-dimensional subband at $B=0$.

We test Equation (\ref{eq:phase condition}) in a device built around an InSb semiconductor nanowire with a superconducting NbTiN contact used to induce superconductivity, and a normal metal Pd contact used to perform tunneling spectroscopy by varying bias voltage $V$ between normal and superconducting contacts (Fig. 1b) (See Methods Summary). Both magnitude and direction of field $B$ can be controlled, as $B$ should be pointed away from the direction of the effective spin-orbit field 
\begin{figure*}[h!]
\centering
  \includegraphics[width=0.9\textwidth]{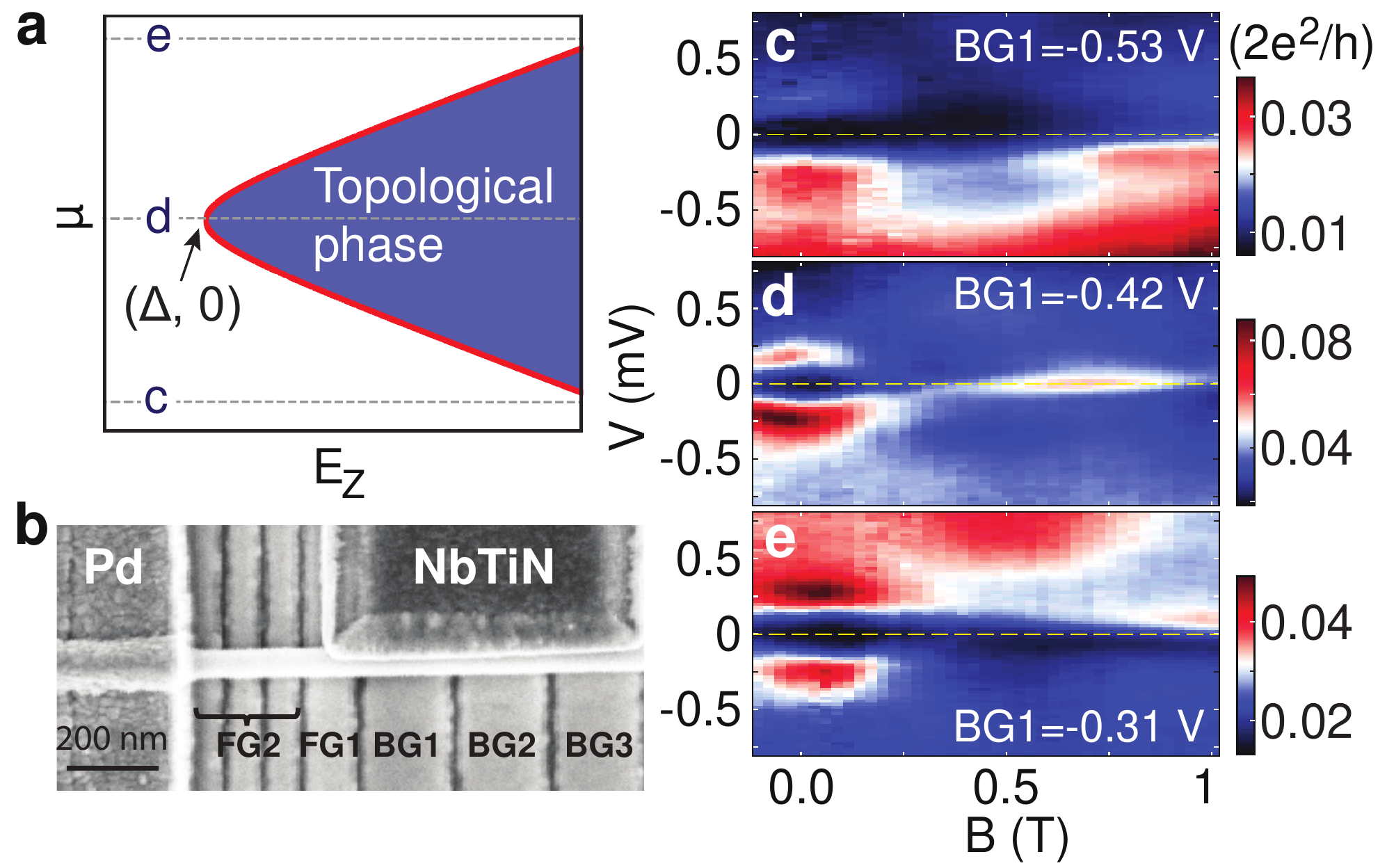}
  \caption{\textbf{Zero-bias peak in a nanowire device controlled by gate voltages.} $\bf{a}$, topological phase diagram described by Eq. (1). Dashed lines indicate settings of $\mu$ in panels e,d and c. $\bf{b}$, scanning electron micrograph of the device used in this work. An InSb nanowire is half-covered by a  superconductor NbTiN, and normal metal Pd contact. The nanowire is placed on FG and BG metal gates. $\bf{c-e}$, differential conductance maps in bias voltage $V$ vs. magnetic field $B$ at three different settings of $BG1$.}
 \label{fig1}
\end{figure*}
 in order to induce MBS. The induced superconducting gap $\Delta$ is set by the NbTiN/InSb interface transparency as well as by the electronic band structure in the nanowire. We treat $\Delta$ as a fixed parameter. Chemical potential $\mu$ in the nanowire is tunable with local gate electrodes placed underneath the nanowire. We adjust voltage on gate $FG1$ to create a tunneling barrier between normal and superconducting sides. Gate $BG1$ located next to the tunneling barrier and underneath the superconductor is used to vary the chemical potential in the nanowire segment under investigation.

We first demonstrate the ability to generate or eliminate a zero-bias peak (ZBP) in conductance over a wide range of $B$ by switching voltage on gate $BG1$. Figs. 1c-1e present scans of bias voltage versus magnetic field applied along the nanowire at three different $BG1$ voltages. The scan obtained at $BG1=-0.42~$V (Fig. 1d) shows a ZBP persistent in magnetic field up to $B=1~$T. When $BG1$ is changed by $\pm 0.11~$V (Figs. 1c,1e), only a gradual closing of the induced gap is observed, with no subgap states up to $1~$T. Thus, Figs. 1c-1e constrain the ZBP phase diagram (horizontal lines in Fig. 1a). The ZBP shows no significant dependence on other gates ($FG2$, $BG2$ and $BG3$), which indicates ZBP is from quantum states located in the nanowire above $BG1$ (see supplementary information).

In Fig. 2 we present the emergence and the evolution of the zero bias peak within the phase space identified in Fig. 1. At zero field, a bias vs. gate scan exhibits an induced gap $\Delta=0.25~$meV (Fig. 2a). We assign conductance maxima at $V=\pm0.25~$mV and around $BG1=-0.4~$V to an increase in the density of states at the bottom of the second one-dimensional subband (see supplementary information for discussion). At $B=0.25~$T (Fig. 2b) the apparent gap decreases but the regime remains qualitatively similar to that at $B=0~$T. We point out that all bias vs. gate data from this device is asymmetric in bias. Namely, resonances that shift to more positive bias voltage with more positive gate voltage dominate. This effect is frequently observed in nanowire devices\cite{MourikScience2012,LeeNatnano2014}, and we attribute this effect to the tunneling barrier asymmetry \cite{KouwenhovenRPP01}.

\begin{figure*}[h!]
\centering
  \includegraphics[width=0.9\textwidth]{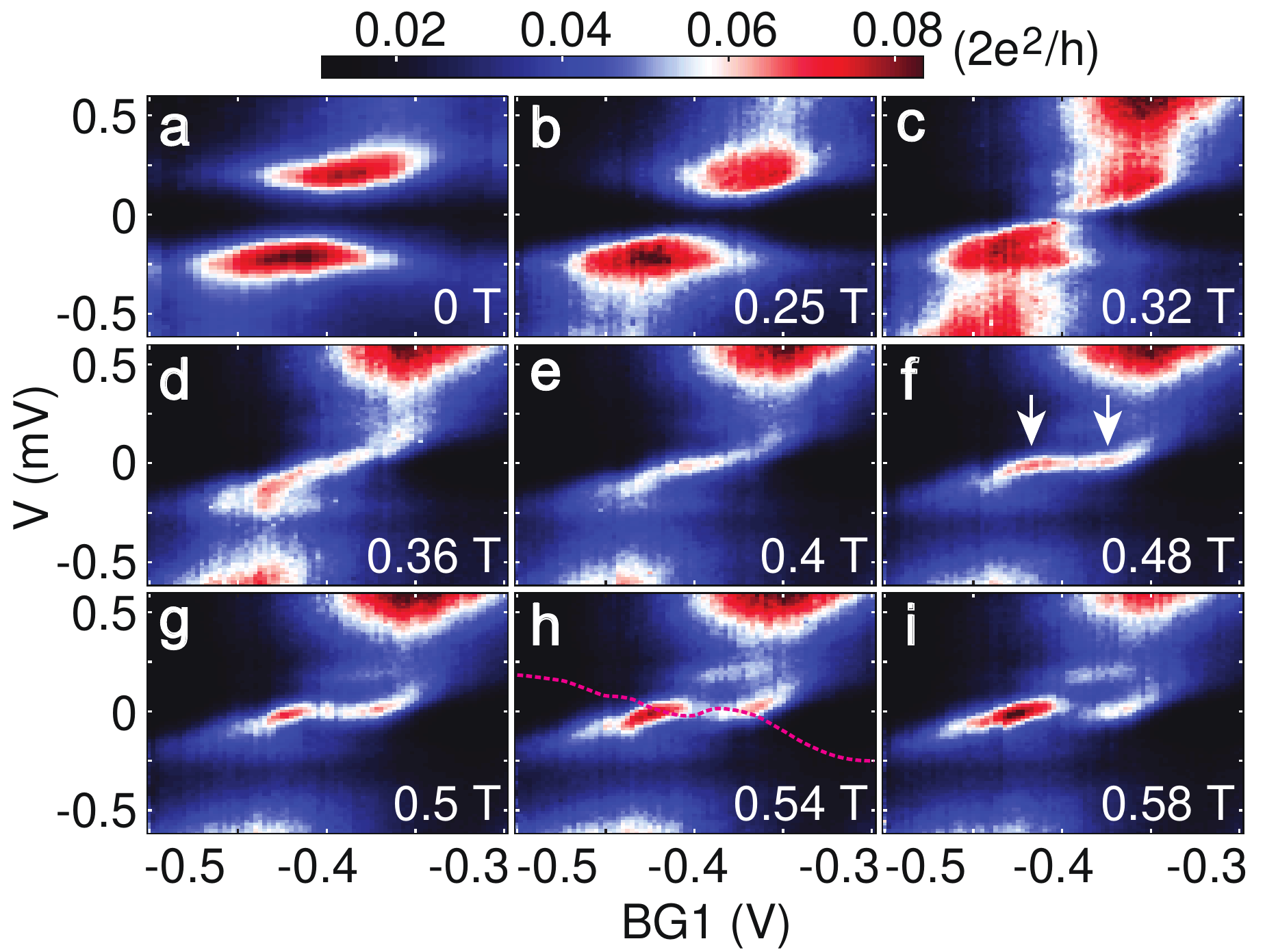}
\caption{\textbf{The emergence of the zero-bias peak.} $\bf{a-i}$, conductance maps in bias voltage $V$ vs. $BG1$ at different magnetic fields indicated in the lower right corner of each panel. Arrows in panel $\bf{f}$ mark the ZBP onset gate voltages plotted in Fig. 3. The dashed line in panel $\bf{h}$ is obtained by tracing the visible maximum in subgap conductance and flipping the resulting trace around $V=0$.
 \label{fig2}}
\end{figure*}

At $B =0.32~$T (Fig. 2c), conductance within the induced gap is increased in the center of the $BG1$ range, giving an indication of a closing gap at $BG1=-0.4~$V. According to theory behind Eq. (1), the gap should close around $\mu=0$ at the topological phase transition. At $B=0.36~$T a well-defined conductance resonance crosses zero bias and extends across the gap (Fig. 2d). The resonance appears to stick to zero bias in a widening range of $BG1$ at higher magnetic fields (Figs. 2e-2f). Towards the edges of each $BG1$ scan, the conductance peak strongly deviates from zero bias and gradually merges into the apparent induced gap. At the boundary defined by Eq. (1), Majorana bound states at the opposite ends of the topological segment of the nanowire grow in length and strongly overlap because of the finite length of the segment. This overlap of the two MBS leads to the MBS energy deviating from zero\cite{AlbrechtNature2016,PradaPRB2012,DasSarmaPRB2012,StanescuPRB2013,RainisPRB2013}.

In addition to the strong deviations from zero bias at the phase boundaries, we observe that for $B\geq 0.5~$T (Figs. 2g-2i) the peak wavers away from zero bias near the center of the scans. Particle-hole symmetry in the superconductor dictates that the energy spectrum within the gap must be symmetric with respect to zero bias. This is not observed due to barrier asymmetry\cite{KouwenhovenRPP01}. However, to propose how the full spectrum inside the gap looks, we trace a subgap resonance in Fig. 2h and flip it along the zero bias line. The full spectrum obtained this way suggests that the small deviations from zero bias also originate from zero-bias peak splitting due to gate-dependent overlap of MBS within the topological phase (see Fig. 4).

We map out the phase diagram of zero-bias peaks: from Fig. 2 and Fig. S3 in supplementary information, we pick the onset points of zero bias peaks in gate $BG1$ as well as in magnetic field, and plot them in Fig. 3. The two data sets obtained this way are consistent with the square root dependence predicted by Eq. (\ref{eq:phase condition}). Based on the diagram, we identify $\mu=0$ at $BG1 =-0.4~$V. The minimal onset field $B =0.33~$T converts into Zeeman energy of $0.4~$meV (using $g=~40$), which is greater than the apparent gap at $B=0~$T.
\begin{figure*}[h!]
\centering
  \includegraphics[width=0.8\textwidth]{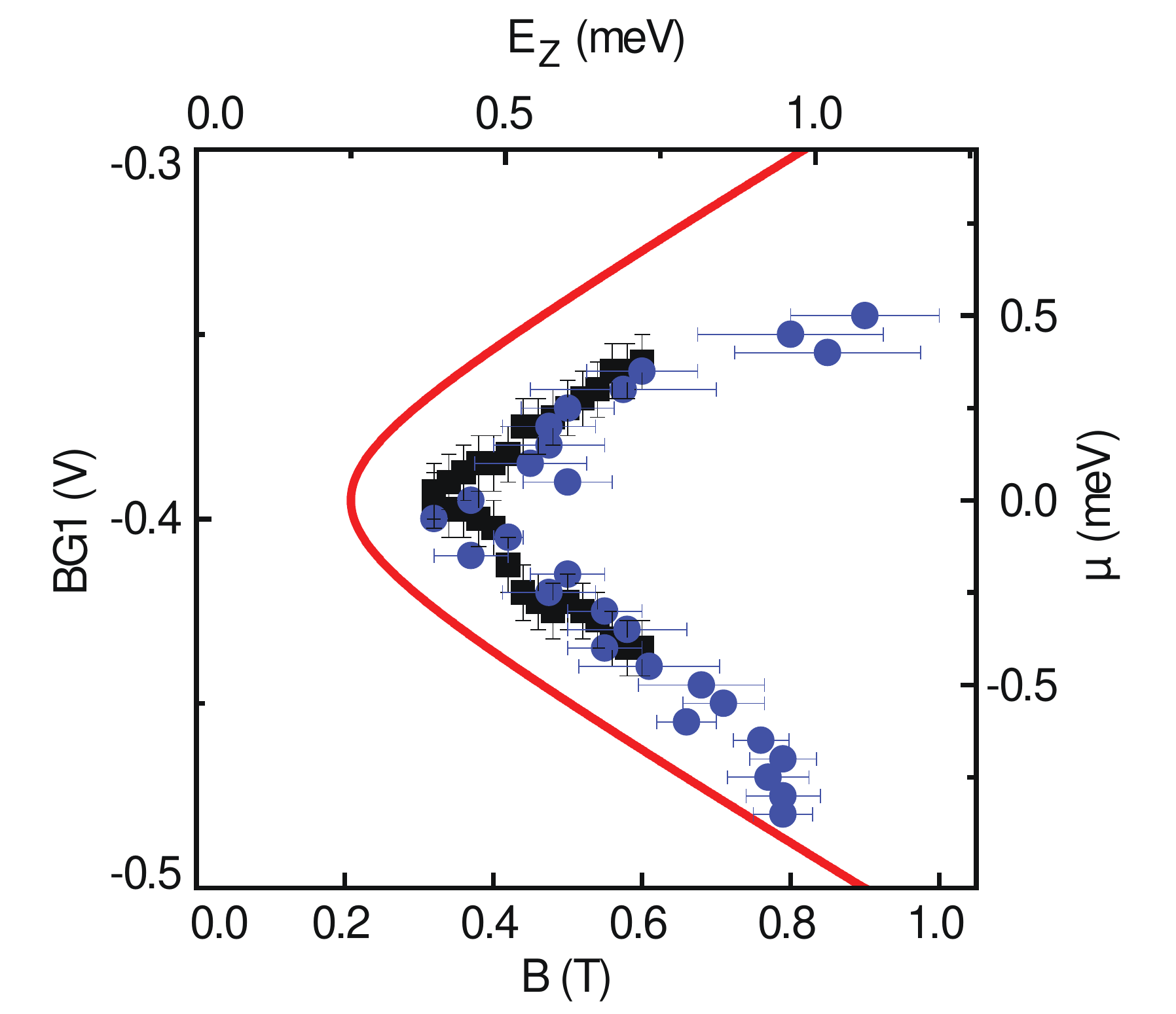}
\caption{\textbf{Phase diagram of zero-bias peaks.} Zero-bias peak onset points are collected from data in Fig. 2 (black squares) and Fig. S3 (blue circles), with error bars judged by deviation of the peak from zero bias within $1/2$ of the full width of half maximun of ZBPs. Data extracted from Fig. S3 are offset by $+0.02~$V in $BG1$ to compensate for a systematic shift due to a charge switch. The top axis $E_{Z}$ is calculated from magnetic field using $g = 40$. The right axis $\mu$ is calculated from $BG1$ according to $10~$meV/V (see Fig. S7a in the supplementary information), and set to be zero at the parabolic vertex, $BG1=-0.395~$V. Equation 1 is plotted in solid line, using $\Delta = 0.25~$mV.
 \label{fig3}}
\end{figure*}
 However, in finite-length superconductors this is expected: due to MBS splitting at the topological transition point the ZBP should onset at a higher field.
For the same reason, ZBP should appear in a narrower range of chemical potential around $\mu=0$ for a fixed field. As a result, the area of the phase diagram with ZBP present is reduced for finite-size systems. In Fig. 3 the theoretical phase transition line predicted by Eq.(1) indeed encircles the extracted ZBP onset points.

The resonance which we investigate as MBS is pinned near zero bias over significant phase diagram area to the right of the onset curve in Fig. 3. The range of ZBP in both chemical potential and Zeeman energy greatly exceeds the ZBP width, which is between $30$ and $100~\mu$eV. The phase diagram area with a ZBP is strongly diminished when magnetic field orientation deviates from the nanowire main axis and approaches the spin-orbit field orientation, previously established as perpendicular to the nanowire (see supplementary information)\cite{Nadj-PergePRL12,MourikScience2012,linPRB12}.

Zero-bias peaks in nanowire devices may also originate from trivial Andreev bound states\cite{LeeNatnano2014}. We present apparent Andreev bound states observed in the same device as well as a similar device at different settings of $BG1$ in supplementary information. In our devices, as opposed to MBS, these states cross zero bias over a narrow range of field and gate voltage that is comparable to the peak width.

Next, we set up a quasi one-dimensional tight-binding model to numerically study a finite-length nanowire under the conditions set by Eq. (1). To match the experimental conditions, a high potential barrier is created above $FG$, potential above $BG1$, $\mu_{BG1}$, is continuously tuned, while potential above $BG2$ and $BG3$ is kept constant (Fig. 4a). Fig. 4a also shows calculated wavefunction amplitude profiles of the two MBS. The left MBS decays into the barrier region above $FG$. The right MBS has an evanescent tail which extends to non-topological regions above $BG2$ and $BG3$. These tails are responsible for a reduced overlap between left and right MBS. Due to the small MBS overlap, the oscillations of the MBS don't reach large amplitudes in energy. Conductance map at zero bias in chemical potential versus Zeeman energy is calculated from the tunneling rates of quantum states (Fig. 4b). The boundary of increased zero-bias conductance is consistent with experimental data in Fig. 3, where the minimum onset field of ZBP is also observed to be larger than $E_Z=\Delta$. The oscillations inside the high conductance region are due to MBS oscillations. 

\begin{figure*}[h!]
\centering
  \includegraphics[width=0.9\textwidth]{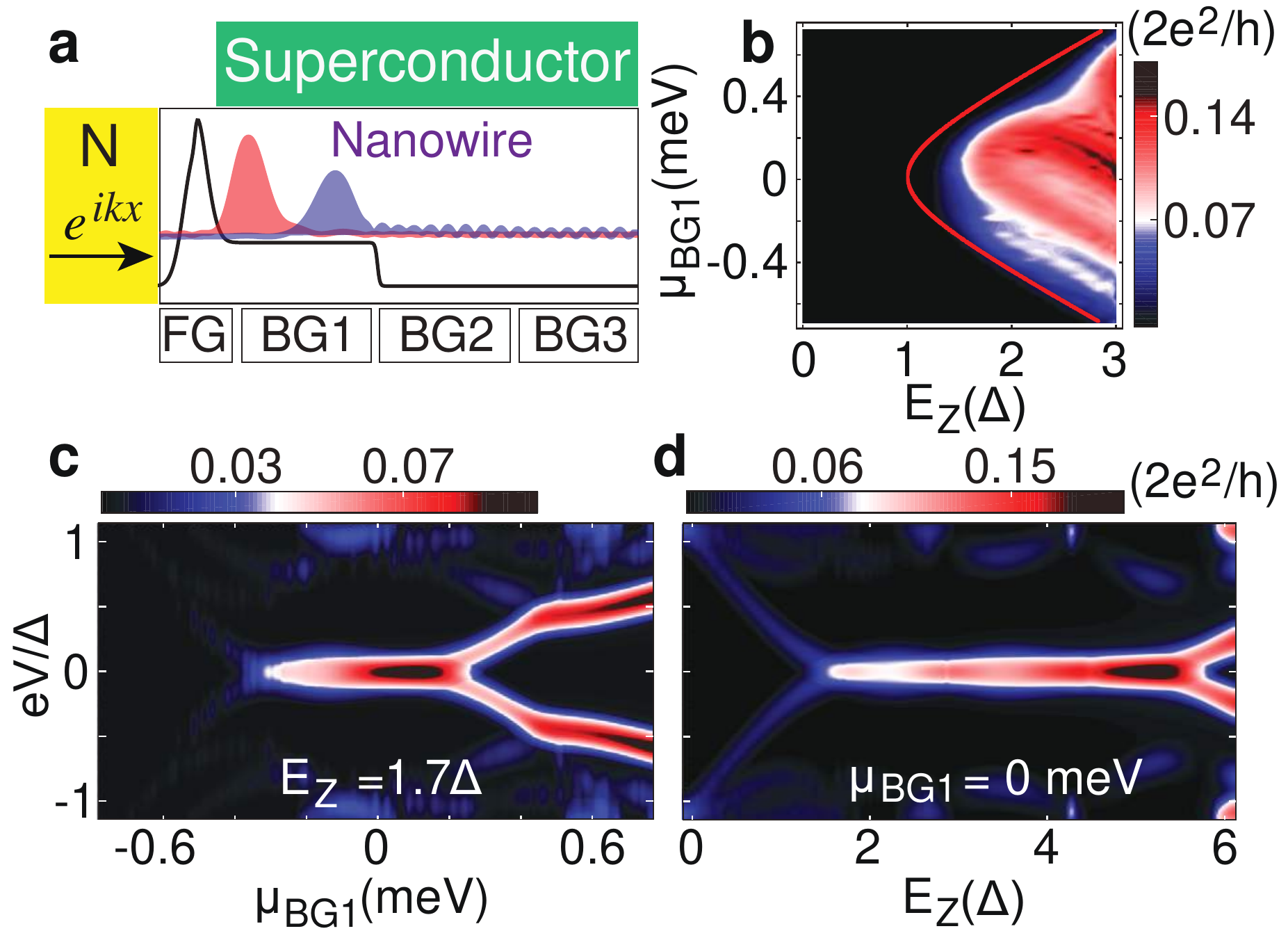}
\caption{\textbf{Tight-binding model results reveal two weakly coupled Majorana bound states.} $\bf{a}$, model schematics. A nanowire is contacted by a superconductor and a normal metal. The potential profile in shown in black curve. A plane wave $e^{ikx}$ coming from $N$ can tunnel into the nanowire through the barrier above $FG$. The chemical potential above $BG1$, $\mu_{BG1}$, is tunable, while potentials above $BG2$ and $BG3$ are fixed. The calculated wavefunction amplitudes for zero-energy states are shown in red and blue. $\bf{b}$, conductance map taken at zero bias. The red curve corresponds to a plot of Eq. (1). $\bf{c}$, conductance map in bias energy vs. chemical potential at $E_{Z}= 1.7~\Delta$. $\bf{d}$, conductance map in bias energy vs. Zeeman energy splitting at $\mu_{BG1}= 0~$meV. In $\bf{b-d}$, thermal broadening is set to $50~\mu$eV to match the experimental ZBP width.
 \label{fig4}}
\end{figure*}

If thermal broadening is included, conductance resonances appear as a single zero-bias peak despite MBS oscillations. In Fig. 4c, at a finite Zeeman splitting of $1.7~\Delta$, we observe an extended ZBP at the center of the map. The zero-bias state occupies a similar range of chemical potential as in the experimental conductance map in Fig. 2f, except that both branches of the spectrum are visible in the simulation. Conductance is suppressed at more negative values of the chemical potential because the states move farther from the probe lead $N$. In the conductance map at chemical potential $\mu_{BG1}=0~$meV (Fig. 4d), an extended zero-bias peak is present from $E_Z = 1.5~\Delta$ up to $E_Z = 5.8~\Delta$. See Methods Summary and supplementary information for calculation details.

Comparison between the model (Figs.4b-4d) and the experiment (Figs.1-3) allows us to conclude that ZBP occurs in the parameter region that is consistent with the predicted topological superconducting phase. This observation makes it significantly less likely that these zero-bias peaks have an origin other than Majorana bound states. Beyond finite-size effects, the detailed experimental phase diagram of zero-bias peaks can be used in future experiments to study how the topological phase is affected by electron-electron interactions\cite{StoudenmirePRB2011}, disorder\cite{adagideliprb14}, vector potentials and electrostatics\cite{NijholtPRB2016,vuiknjp2016}.

\begin{methods}
InSb nanowires are grown by Au-catalyzed Vapor-Liquid-Solid mechanism in a metalorganic vapor phase epitaxy reactor. Nanowires are deposited onto bottom gate chips using a micro-manipulator. Nanowires have a diameter between 60-100~nm. The bottom gates are made of Ti(5~nm)/Au(10~nm), with FG gates 50/100~nm wide and BG gates 200~nm wide. A layer of high-$\kappa$ dielectric HfO$_{2}$(10~nm) is deposited onto the bottom gates. Prior to contact deposition, the nanowire is processed in a 1/500 diluted ammonium sulfide solution  by baking at 55$~^{\circ}$C for 30 minutes to remove a native oxide layer. The superconducting contact is a trilayer of Ti(5~nm)/NbTi(5~nm)/NbTiN(180~nm) optimized to suppress subgap conductance\cite{ZhangArxiv2016}. While the induced gap is clearly visible in a wide gate range, the subgap conductance remains finite and the gap is increasingly soft at finite field (see Fig. S1 in supplementary information). Coverage of the nanowire by the superconductor is reduced to minimize gate screening and enlarge the range of tunable chemical potential. The normal contact is a Ti(15~nm)/Pd(150~nm) stack, before its deposition a gentle argon plasma cleaning is performed $in$ $situ$.  

Measurements are performed in a dilution refrigerator at a base temperature of 30~mK, by standard low-frequency lock-in technique (77.77~Hz, 5$~\mu$V). Multiple stages of filtering are used to enhance signal-to-noise ratio. For all the measurements, bias voltage is applied to the normal contact and the superconducting contact is grounded.
 
The theoretical model consists of a tight binding Hamiltonian for both the lead and the nanowire with induced superconductivity. The Hamiltonian includes hopping and chemical potential in both the nanowire and the metallic lead. In addition, in the nanowrie there is Rashba spin-orbit coupling, a magnetic field oriented along the wire, induced superconductivity, and potentials above each gate. The differential conductance is then extracted by applying plane wave boundary conditions at the end of the lead and solving for the anomalous reflection coefficients.  
\end{methods}

\bibliographystyle{unsrt}
\bibliography{Ref_TSC.bib}

\begin{addendum}
 \item We thank A. Akhmerov, S. De Franceschi, V. Mourik, F. von Oppen, D. Pekker, F. Pientka, M. Wimmer for valuable discussions. Work is supported by NSF DMR-125296 and ONR N00014-16-1-2270. T.S. acknowledges support from NSF DMR-1414683.
 \item[Author Contributions] D.C., S.P. and E.B. grew InSb nanowires. J.C., P.Y. and M.H. fabricated devices. J.C., P.Y. and S.F. performed the measurements. J.S. and T.S. performed numerical simulations. All authors analyzed results and wrote the manuscript.
 \item[Correspondence] Correspondence and requests for materials should be addressed to S.M.F.~(email: frolovsm@pitt.edu).
\end{addendum}

\newpage

\setcounter{figure}{0}
\setcounter{equation}{0}
\setcounter{section}{0}

\begin{center}
{\Large\textbf{Supplementary Information}}
\end{center}

\tableofcontents

\bigskip

{\centering \section{Dependence of ZBP on gates, magnetic field and field orientation}
}

We first present linecuts from Figs. 1 and 2 in the main text, shown in Fig. \ref{Fig.S1}. At zero field, we observe an induced gap of $\Delta=~0.25~$mV, and a ratio of conductance outside and inside the gap of $2.8$ (Fig. \ref{Fig.S1}a). Figs. \ref{Fig.S1}b-f are linecuts of Figs.2a, 2f and Figs. 1c-e, respectively. 

We then demonstrate dependence of the ZBP on other gates besides $BG1$. We set magnetic field $B=0.5~$T and gate $BG1=-0.42~$V, where a pronounced ZBP appears, then scan the other gates individually, as shown in Fig. \ref{Fig.S2}. ZBP does not move with gates $FG1$, $FG2$, $BG2$ and $BG3$. Given the fact that the ZBP is only tunable with $BG1$, we conclude that the quantum states giving rise to ZBP must be located within the nanowire segment above $BG1$.

We now turn to scans of bias voltage versus magnetic field in small steps of gate $BG1$, shown in Fig. \ref{Fig.S3}. All scans demonstrate a conductance resonance originating from the zero-field gap edge and evolving into a zero-bias peak at finite magnetic field. The onset points of zero-bias peaks are marked with arrows. The ZBP onset field $B_{onset}$ is strongly dependent on $BG1$. For BG1=$-0.46~$V, $B_{onset}=0.8~$T, while for $BG1=-0.395~$V, $B_{onset}=0.33~$T. The onset point first shifts to lower fields as $BG1$ is increased (from $-0.475~$V to $-0.395~$V), and then shifts to higher fields (from $-0.395~$V to $-0.33~$V). Past the onset point, the ZBP or a split peak persist near zero bias for a significant range of magnetic field. 

The onset points from scans in Fig. \ref{Fig.S3} have been picked and plotted in magnetic field vs. $BG1$ in Fig. 3 in the main text, which demonstrates the phase diagram of ZBP. 

In Fig. 2 of the main text, we have demonstrated that ZBP  is tunable with $BG1$ for fields up to $B=0.58~$T. Figs. \ref{Fig.S4}a-d are similar to Fig. 2. However, at $B=0.6~$T (Fig. \ref{Fig.S4}e), the region with ZBP becomes fragmented into two parts. With magnetic field increasing further, the left part shows additional fragmentation and loops can be resolved (Figs. \ref{Fig.S4}f-h). In the coupled Majorana interpretation these are manifestations of several oscillation periods\cite{DasSarmaPRB2012,PradaPRB2012,StanescuPRB2013,RainisPRB2013}. It is also possible that this is a manifestation of orbital effects that are expected to play a larger role at higher fields\cite{NijholtPRB2016}. However, because only one branch of the spectrum is clearly visible it is not possible to make a more detailed analysis.

Next, we rotate the in-plane magnetic field, as shown in Fig. \ref{Fig.S5}. From zero degree to an angle of $\pi/6$ with respect to the nanowire main axis, the zero-bias peak is present at fields above $0.3~$T (Figs. \ref{Fig.S5}a-c). At larger angles from $\pi/4$ to $\pi/2$, no extended ZBP is observed (Figs. \ref{Fig.S5}d-g). From angles $5\pi/6$ to $\pi$, the ZBP restores again(Figs. \ref{Fig.S5}h-j). In fact, no clear superconducting gap or subgap states are observed above $0.3~$T for larger angles in Fig. \ref{Fig.S5}. Fig. \ref{Fig.S5}k gives an overview of the angle dependence of the ZBP. At a fixed magnetic field $B=0.5~$T, ZBP is present in a window of angles between $-0.2\pi$ and $0.2\pi$, at angles above $0.2 \pi$ the peak deviates from zero bias. Since the direction of spin-orbit effective field was previously established to point at $\pi/2$, the angle dependence of ZBP indicates the relevance of spin-orbit interaction in making the subgap state survive to large Zeeman splittings. These observations are in line with previous reports on ZBP in hybrid devices, as well as with theoretical calculations\cite{MourikScience2012,LinPRB12}.

In Fig. \ref{Fig.S6} we present gate scans at different magnetic fields, while keeping the field direction perpendicular to the nanowire. Similar to scans at small angles, it shows that conductance increases with magnetic field in the center of the gate range, then a conductance resonance crosses zero bias and extends across the bias range. However, at this angle perpendicular to the nanowire there is no ZBP over an extended range of $BG1$. At higher fields, the resonance becomes blurred and possibly splits.

{\centering \section{Conductance resonances above the gap}
}

We present zero-field data in the expanded range of $BG1$ and $V$ in Fig. \ref{Fig.S7}. Note that the gate voltages do not perfectly match the main text due to a charge jump from a scan of large range of $BG1$. The regime presented in Figs. 1-2 is now in the vicinity of $BG1 = -0.5 ~$V. We observe a broad conductance resonance that crosses the gap (marked by the solid line in Fig. \ref{Fig.S7}a). This is the resonance that coincides with the ZBP presented in the main text. Other resonances above the gap are discernible at more positive gate voltages. While all of these resonances disperse strongly with $BG1$, they have a relatively weak dependence on $FG1$ (Fig. \ref{Fig.S7}c), and don’t have measurable dependence on $FG2$ (Fig. \ref{Fig.S7}d). Indeed, the positions of the resonances are greatly affected by $BG1$, but not by other gates. Note that the presented range of $FG1$ is smaller, but because this gate is not fully under the superconductor thus it is much stronger electrostatically coupled to the nanowire than $BG1$. The relative insensitivity to other gates suggests that quantum states giving rise to resonances in Fig. \ref{Fig.S7}a are localized in the nanowire above $BG1$. The resonances at $BG1 < 0.25~$V are not sensitive to $BG2$, while for $BG1>0.25~$V we observe some more faint resonances that are tunable with both $BG1$ and $BG2$ (Fig. \ref{Fig.S7}e). Thus, when $BG1$ tunes the nanowire to much higher density, states that extend across several gates underneath the superconductor become resolved. This regime has not been further studied in this device and is a topic of a future study. 

One explanation for the resonance at $BG1 =-0.5~$V is that it is a manifestation of the density of states singularity near the edge of a 1D subband. This is indirectly supported by pinch-off traces of $BG1$, obtained with barrier FG1 open (in circle) and closed (in square) (Fig. \ref{Fig.S7}b). $BG1$ is not capable of completely stopping current through the device because it is shunted by a high transparency contact to the superconductor. However, a transition between two current values which we interpret as zero density and high density above $BG1$ is observed in the vicinity of $BG1 = -0.5~$V when $FG1$ is in the transmitting regime (open), indicating that this gate voltage is within the pinch-off region of $BG1$, plausibly near one of the subband edges.

In a multiband quantum wire, a topological phase onset similar to that described by Eq. (1) occurs close to the chemical potential at which each new one-dimensional subband aligns with the Fermi level. The subband spacing in InSb nanowires was previously found to be $10-15~$meV\cite{vanweperennanolett13,Kammhubernanolett16}. Thus for fields $B =1~$T, or $E_{Z} = 1.2-1.5~$meV, gate voltage ranges with MBS should be separated by much larger gate voltage ranges without MBS.

The second resonance (marked by the dashed line in Fig. \ref{Fig.S7}a) is separated from the first one by $2~$meV, an energy calculated from the dispersion of the resonances in bias and gate. This is a spacing smaller than the typical subband spacing. Thus, at least not all of the resonances are due to the subband edge singularities – this is also supported by the fact that the first and second resonances disperse differently with BG1 suggesting that they correspond to spatially distinct regions with different capacitive coupling to $BG1$. The magnetic field evolution of the resonances that extend into the normal state and the deeper discussion of their origins will be published separately.

Our analysis concludes that the resonance at $BG1 = -0.5~$V corresponds to the second subband edge (labeled by $N=2$ in Fig. \ref{Fig.S7}c). It was not possible to clearly isolate another broad resonance similar to that at $BG1 = -0.5~$V, and study another candidate MBS in this device. We hypothesize that the next such resonance is located near $BG1 = 0.1~$V (labeled by $N=3$ in Fig. \ref{Fig.S7}c), where too many trivial resonances appear simultaneously with a broader conductance feature of slope similar to the dashed line in Fig. \ref{Fig.S7}a. This region is separated from the region marked by the solid line by approximately 10 meV in chemical potential. The first subband may lie at more negative gate voltages near $BG1= -1.2~$V. A broad resonance can be resolved in that range of $BG1$ in Fig. \ref{Fig.S7}c(labeled by $N=1$), at $FG1=0.17~$V. However, the conductance in that regime was too low to perform a thorough study. The fact that we did not reach the last subband is confirmed by the fact that induced gap is observed in Fig. \ref{Fig.S7}a (marked by arrows) at voltages more negative than the resonance marked by the solid line. Indeed, if this were the first subband then we would not expect any induced superconductivity at negative chemical potentials where the density in the semiconductor would be zero.

{\centering \section{Trivial Andreev bound states}
}

In Fig. \ref{Fig.S8} we study the subgap spectrum in the vicinity of the second resonance (marked by the dashed line in Fig. \ref{Fig.S7}a). Here two subgap resonances appear within the gap at finite field. However, these resonances cross zero bias over a much narrower field interval compared to states studied in the main text. Overall, their field and gate dependence is highly reminiscent of Andreev bound states previously reported in nanowire quantum dots coupled to superconductors\cite{LeeNatnano2014}. We thus conclude that these states are not MBS but rather that they are trivial states localized near $FG1$ above $BG1$. We note that such resonances are also expected in finite-length topological superconductors at chemical potentials much larger than zero (see theory discussion below and Fig. \ref{TS5}).

Next, we present data on trivial Andreev bound states (ABS) in another device of very similar geometry, as shown in Fig. \ref{Fig.S9}. In this device an accidental quantum dot formed above $BG1$, and Coulomb blockade was observed (data not shown). Also in this device the data is less asymmetric, and both branches of the spectrum are clearly resolved. Figs. \ref{Fig.S9}a-c demonstrate Andreev bound states in $BG1$ at different magnetic fields. We see a gap feature similar to the one presented in the main text at zero field. At $B=0.25~$T, two conductance resonances cross at zero bias (Fig. \ref{Fig.S9}b). At higher field a characteristic loop is formed in the center of the scan (Fig. \ref{Fig.S9}c). Figs. \ref{Fig.S9}d-j demonstrate Andreev bound state evolution with magnetic field for different settings of $BG1$. A general feature is that a pair of conductance resonances split with magnetic field, and the inner branches cross at zero bias, while the outer branches merge into the gap edge. We point out that at zero field the energies of ABS are different at different $BG1$, thus they reach zero bias and cross their opposite bias copy at different fields (having similar Zeeman splitting). 

The ZBP observed here is from two Andreev levels crossing, it does not pin to zero bias for any significant field range. The crossing point is robust in field angle, i.e. it occurs at all tested field angles though it shifts position in field according to the g-factor anisotropy\cite{LeeNatnano2014}. Approximately, the width of a single Andreev bound state in bias voltage is $0.1~$meV, and the extent of ZBP in field is $0.1~$T which is $0.1-0.15~$meV in Zeeman splitting (Fig. \ref{Fig.S9}h). Thus the resonances don't pin to zero bias for a range of energies much longer than the peak width. An exception is shown in Fig. \ref{Fig.S9}d, where a longer ZBP is observed in a narrow range of gate voltage. This extended ZBP is due to additional ABS approaching zero bias at $B > 0.5 T$. We also note that the ABS zero-bias crossing point would trace out a curve very similar to that in Fig. 3 of the main text, if plotting in the phase diagram space of $E_Z$ vs $\mu$. However, the region separated by that curve would have  zero conductance at zero bias except for the points right at the ABS zero crossing.

In summary, tunneling measurements of trivial ABS and topological MBS share many features. Both types of states evolve from an apparent induced gap, and reach zero bias at finite field. Both types of states can produce a zero-bias peak onset curve consistent with Equation (1). Thus, extreme caution should be used in all experiments when identifying non-trivial topological states. In this work, we differentiate the two types of bound states through the extent of zero bias peaks in field and gate range, field orientation dependence and through comparison with theory. We find that when magnetic field is aligned with the nanowire, the ZBP studied in the main text is pinned to zero bias over a range of Zeeman splitting and chemical potential that greatly exceeds the ZBP width in bias, thus filling up the inner area of the phase diagram with zero or near-zero energy states. This is not the case when the field is not aligned with the nanowire, or for other ZBPs which we identify as trivial.

{\centering \section{Description of the numerical model}
}

The theoretical modeling of the normal metal-semiconductor-superconductor hybrid system is based on a simple quasi one-dimensional tight-binding model consisting of $N_y$ parallel coupled chains (Fig. \ref{TS1}). The metallic lead is described by the Hamiltonian
\begin{myequation}
H_M=-\sum_{{\bm i},{\bm \delta}} t_{m}^\delta (c^{\dagger}_{\bm i}c_{{\bm i}+{\bm \delta}}+c^{\dagger}_{{\bm i}+{\bm \delta}}c_{\bm i})+\mu_m\sum_{i}c^{\dagger}_{\bm i}c_{\bm i},
\label{EqThS1}
\end{myequation}
where ${\bm i}=(i_x, i_y)$, with $1\leq i_y \leq N_y$ labeling the chains and $0\leq i_x \leq N_m$ labeling the position along the chains, while ${\bm \delta}=(\delta_x, \delta_y)$ designates nearest-neighbors along and across the chains. The hopping parameters along and across the chains are $t_m^{\delta_x} = t_m = 3.8~$meV, $t_m^{\delta_y} = t_m^\prime = 3.8~$meV and the chemical potential of the normal lead is $\mu_m = -7.6~$meV. The electron creation operator is written in spinor form, $c_{\bm i}^\dagger = ( c_{{\bm i}\uparrow}^\dagger, c_{{\bm i}\downarrow}^\dagger)$.

The semiconductor wire, including the effects of gate potentials, spin-orbit coupling, applied Zeeman field, and proximity-induced superconductivity, is modeled by the tight-binding Hamiltonian
\begin{myequation}
\begin{aligned}
H_{SM} =& -\sum_{{\bm i},{\bm \delta}} t_{sm}^\delta (c^{\dagger}_{\bm i}c_{{\bm i}+{\bm \delta}}+c^{\dagger}_{{\bm i}+{\bm \delta}}c_{\bm i})+\sum_{i}(\mu_{sm} +V_{\bm i})c^{\dagger}_{\bm i}c_{\bm i} 
\\
&+ \frac{i}{2}\sum_{{\bm i},{\bm \delta}}\alpha_R^\delta(c^{\dagger}_{{\bm i}+\delta_x}\hat{\sigma}_y c_{{\bm i}}-c^{\dagger}_{{\bm i}+{\delta_y}}\hat{\sigma}_x c_{\bm i} + h.c.)  \\
&+ \Gamma\sum_{\bm i}c^{\dagger}_{\bm i} \hat{\sigma}_x c_{\bm i} + \Delta \sum_{\bm i}(c^{\dagger}_{{\bm i}\uparrow}c^{\dagger}_{{\bm i}\downarrow}+c_{{\bm i}\uparrow}c_{{\bm i}\downarrow}), 
\end{aligned}
\end{myequation}
where $\hat{\sigma}_\mu$, with $\mu=x, y, z$, are Pauli matrices. The position along the chains containing $N_{sm}$ sites is labeled by $i_x$, with $N_m+1\leq i_x \leq N_m+N_{sm}\equiv N$.
The matrix elements for hopping along and across the chains are 
$t_{sm}^{\delta_x}=t_0=9.5~$meV and $t_{sm}^{\delta_y}=t_0^\prime=1.1~$meV, respectively, and the chemical potential of the wire is $\mu_{sm}=-5.2~$meV. 
The position-dependent local term $V_{\bm i}=V(i_x)$ describes the potential generated by the bottom gates as well as tunnel barrier potential. 
In our model we separate the $V_{\bm i}$ into three regions, the barrier potential located at the left side of the wire, region above $BG1$ and runs over forty sites, then region above $BG2, BG3$ along the remainder of the wire and remains at zero for all calculations. The strengths of the longitudinal and transverse components of the Rashba spin-orbit coupling are characterized by the coefficients $\alpha_R^{\delta_x}=\alpha_{R}=0.2~$meV and $\alpha_R^{\delta_y}=\alpha_R^\prime=0.7~$meV, respectively, while the Zeeman splitting corresponding to a magnetic field applied along the wire is given by $\Gamma$\cite{Vanweperenprb15}. Proximity-induced superconductivity is described by the last term in Eq. (S2), with $\Delta = 0.25~$meV representing the induced pair potential. The coupling between the normal lead and the proximitized semiconductor wire is described by the coupling Hamiltonian with  $\tilde{t}=2.3~$meV the hoping energy between the lead and the semiconductor.
\begin{myequation}
H_{M-SM}=\tilde{t}\sum_{i_y} (c_{N_m i_y}^\dagger c_{N_m+1 i_y} c_{N_m+1 i_y}^\dagger c_{N_m i_y}).
\label{EqThS3}
\end{myequation}

To calculate the differential conductance for charge tunneling from the metallic lead into the end of the semiconductor wire we use the Blonder-Tinkham-Klawijk (BTK) formalism \cite{BlonderPhysRevB1982}. More specifically, we calculate the reflection and
transmission coefficients by solving the Bogoliubov-de Gennes (BdG) equation for $H=H_M+H_{SM}+H_{M-SM}$ with open boundary conditions on the normal lead. In the experiment the bulk superconductor is grounded and we assume that current propagates trough it as supercurrent. Consequently, the transmission coefficient vanishes and we only have to account for the normal and Andreev reflection.  We note that in the presence of a quasiparticle current (e.g., at high bias voltage) the bulk superconductor has to be explicitly included in the formalism, with appropriate boundary conditions. The BdG equation that determines the normal ($r_N$) and Andreev ($r_A$) reflection coefficients is 
\begin{myequation} \label{eq:bound}
\begin{split}
\sum_{j_x=0}^{N}\sum_{j_y=1}^{Ny}\sum_{\sigma^\prime} ({\cal H}_{{\bm i}\sigma,{\bm j}\sigma'}-\omega~\! \delta_{{\bm i},{\bm j}}\delta_{\sigma,\sigma'})\Psi_{{\bm j},\sigma^\prime}=0, 
\\
~ {\rm for}~ i_x=1,\dots,N ~~i_y=1,\dots,N_y ~~{\rm and} ~\sigma=\pm,
\end{split} 
\end{myequation}
where ${\cal H}$ is the (first quantized) BdG Hamiltonian that can be easily extracted from Eqs. (S1-S3) by writing the total (second quantized)  Hamiltonian in the form $H = \frac{1}{2}\sum_{{\bm i}, {\bm j}} \psi_{\bm i}^\dagger {\cal H}_{{\bm i}{\bm j}}\psi_{\bm j}$, where the fermion creation and annihilation operators are contained in the Nambu spinors,   $\psi_{\bm i}^\dagger = (c_{{\bm i}\uparrow}^\dagger   c_{{\bm i}\downarrow}^\dagger c_{{\bm i}\uparrow}   c_{{\bm i}\downarrow})$. Eq. (S4) has to be solved for all chains, $1\leq i_y\leq N_y$. For the normal lead it is convenient to define the transverse modes $\phi_\nu$ ($1\leq \nu \leq N_y$) characterized by the wave functions $\phi_\nu(i_y) = \sqrt{2/(N_y+1)}\sin[i_y\nu \pi/(N_y+1)]$. Each pair $(\nu,\sigma)$ corresponding to a given transverse mode and spin orientation defines a transport channel and the reflection coefficients $r_N$ and $r_A$ are matrices with matrix elements indexed by these channel labels. 

The boundary conditions for an incoming electron in channel $(\nu,\sigma)$ can be expressed in terms of reflection coefficients as
\begin{myequation}
\begin{aligned}
\Psi_{j_x=0,j_y} &=\phi_\nu(j_y)
\begin{pmatrix} \delta_{\sigma,\uparrow}\\ \delta_{\sigma_\downarrow}\\0\\0 \end{pmatrix}
+ \sum_{\nu^\prime}\phi_{\nu^\prime}(j_y)
\begin{pmatrix}
[r_N]_{\nu\sigma,\nu^\prime\uparrow}\\ 
[r_N]_{\nu\sigma,\nu^\prime\downarrow}\\ 
[r_A]_{\nu\sigma,\nu^\prime\uparrow}\\ 
[r_A]_{\nu\sigma,\nu^\prime\downarrow}
\end{pmatrix}
\\
\\
\Psi_{j_x=1,j_y} =\phi_\nu(j_y)\begin{pmatrix} \delta_{\sigma,\uparrow}\\ \delta_{\sigma_\downarrow}\\0\\0 \end{pmatrix} e^{i k_e^\nu a}
 &+ \sum_{\nu^\prime}\phi_{\nu^\prime}(j_y)\begin{pmatrix}
[r_N]_{\nu\sigma,\nu^\prime\uparrow}\\ 
[r_N]_{\nu\sigma,\nu^\prime\downarrow}\\ 
0\\ 
0 \end{pmatrix}e^{-i k_e^{\nu^\prime} a} 
+ \sum_{\nu^\prime}\phi_{\nu^\prime}(j_y)\begin{pmatrix}
0\\ 
0\\ 
[r_A]_{\nu\sigma,\nu^\prime\uparrow}\\ 
[r_A]_{\nu\sigma,\nu^\prime\downarrow} \end{pmatrix} e^{i k_h^{\nu^\prime} a},  
\end{aligned}
\end{myequation}
where $a=0.1~$nm is the lattice constant while $k_e^\nu$ and $k_h^\nu$ are wave vectors in the lead at energies $\omega$ and $-\omega$, respectively, 
\begin{myequation}
k_{e(h)}^\nu(\omega) = \cos^{-1}\left(-\frac{\mu_m +\epsilon_\nu\pm\omega}{2t_m}\right),
\end{myequation}
where $\epsilon_\nu = 2t_m^\prime \cos[\nu \pi/(N_y+1)]$.

With the boundary conditions given by Eq. (S5), the BdG equation (S4) reduces to a system of $N$ linear equations (with $N$ unknown coefficients) from which one can easily extract the reflection coefficients $[r_{N(A)}]_{\nu\sigma,\nu^\prime\sigma^\prime}$ for a given incoming channel $(\nu \sigma)$ and all possible reflection channels $(\nu^\prime \sigma^\prime)$. The procedure is repeated by $2N_y$ times, once for each incoming channel. Since normal reflection does not contribute to the injected current, the conductance is given by the Andreev reflection coefficients \cite{BlonderPhysRevB1982}. In the unit of $2e^2/h$ we have
\begin{myequation}
\frac{dI}{dV}= \sum_{\nu,\nu^\prime}\sum_{\sigma,\sigma^\prime} |[r_A(\omega=V)]_{\nu\sigma, \nu^\prime\sigma^\prime}|^2,
\end{myequation}
where $V$ is the bias voltage. To include the effect of finite temperature, the conductance is broadened by convolving with the Fermi function, 
\begin{myequation}
G(V,T) =\int d\omega \frac{G_0(\omega)}{4 T \cosh[(V-\omega)/2k_bT]},
\end{myequation}
where $G_0=dI/dV$ is the zero temperature conductance given by Eq. (S7) at voltage bias $\omega$.

{\centering \section{Supplementary numerical results}
}

In the presence of a nonuniform, step-like gate potential (see main text, Fig. 4a) the low-energy states have most of their weight either in the wire segment above $BG1$, or in the segment above $BG2$ and $BG3$. The states that contribute to the measured differential conductance are almost entirely confined to the region above $BG1$, as experimentally confirmed by the insensitivity of the measured $dI/dV$ to variations of the $BG2-BG3$ potential. 
Therefore, it is tempting to model the system as a short wire with a length equal to the $BG1$ segment \cite{StanescuPRB2013}. Indeed, the lowest energy state corresponding to a step-like potential is very similar to the lowest energy state of a short wire, as shown in Fig. \ref{TS2}. Note that, in both cases the lowest energy state can be viewed as a pair of overlapping Majorana bound states. However, in the case of a long wire with a step-like potential the rightmost Majorana can leak into the $BG2-BG3$ region, which results in significantly reducing the overlap with the other Majorana bound state as compared to the short wire case. Consequently, the amplitude of the energy splitting oscillations for a long wire with step-like potential will differ significantly from that corresponding to a short wire, as illustrated in the lower panels of Fig. \ref{TS2}. In particular we emphasize the qualitatively different dependence of this amplitude on the Zeeman field: for a short wire the amplitude of the oscillations increases rapidly  with the Zeeman field \cite{StanescuPRB2013}; by contrast, for the long wire with step-like potential the amplitude decreases with the field (within a certain range of physically relevant values of the magnetic field). 

Having established that a step-like background potential generates features that are qualitatively different from those corresponding to a short wire, we now focus on understanding the dependence of the low-energy spectrum on the $BG1$ gate potential and the emergence of nearly-zero energy states at finite magnetic fields. The dependence of the low energy spectrum on $\mu_{BG1}$ is shown in the left panels of Fig. \ref{TS3} for three different values of the Zeeman field. We also calculate the differential conductance for tunneling into the (left) end of the wire. The corresponding conductance maps are shown in the right panels of Fig. \ref{TS3}. 

At zero field all states are above the induced superconducting gap. Note that each parabola-like curve corresponds to a low-energy state that ``lives'' in the region above $BG1$ and changes its energy as $\mu_{BG1}$ varies. For $\mu_{BG1}<-5.5~$meV these states have energies way above the chemical potential and do not contribute to the low-energy spectrum. Increasing  $\mu_{BG1}$ results in the chemical potential reaching the bottom of a confinement-induced band, then successively crossing the low-energy states from this band, which results in the ``parabolas'' shown in Figs. \ref{TS3}a, b. Note that, the strength of the signature of each state in conductance map depends on its amplitude at the tunnel barrier. For the parameters in this calculation, this strength increases as energy goes up, the weakest signature corresponding to the bottom of the band (see Fig. \ref{TS3}b).

At finite magnetic field, the lowest energy state approaches zero within a finite range of gate voltages corresponding to the chemical potential near the bottom of the band, e.g., $-5.1~$meV$ \leq \mu_{BG1}\leq -4.6~$meV for $E_{Z} = 1.7~\Delta$. 

Next, we fix the gate voltage and calculate the dependence of the energy spectrum on the Zeeman field. The results for $\mu_{BG1}=-4.7~$meV (corresponding to the zero-energy crossing in Fig. \ref{TS3}e, f) are shown in Fig. \ref{TS4}. For $E_{Z} > 1.7~\Delta$ the lowest energy state is pinned near zero for values of the Zeeman field within a range of $4\delta=1~$meV.  

However, for larger values of $\mu_{BG1}$, i.e., when the chemical potential is not in the vicinity of the bottom of the band, the lowest-energy state does not pin to zero, although it does cross zero at certain (large) values of the Zeeman field. This situation is illustrated in Fig. \ref{TS5}. We note that in this regime the zero-energy crossings represent a generic feature but the tendency of pinning to zero energy is absent. Qualitatively, this behavior can be understood in terms of overlapping Majorana bound states with different characteristic length scales $\xi_M$. Roughly, $\xi_M$ is proportional to the Fermi velocity and, therefore, is minimal when the chemical potential is close to the bottom of the band and increases as we move up in energy. The case illustrated in Fig. \ref{TS4} corresponds to $\xi_M$ being smaller than the length of the $BG1$ region, which can accommodate two Majorana bound states without significant overlap resulting in a low-energy mode that is pinned to zero. By contrast, the regime illustrated in Fig. \ref{TS5} corresponds to $\xi_M$ comparable to or larger than the length of the $BG1$ segment, which results in a strong overlap of the Majoranas. A pair of such strongly overlapping Majorana modes can be more conveniently viewed as a (regular) Bogoliubov quasiparticle that, generically, has finite energy but may have zero energy for a discrete set of parameters (i.e., values of the gate potential and Zeeman field). 

These results, which are obtained using a rather simple model of the nanostructure, are qualitatively consistent with the experimental findings. Of course, the real structure is characterized by additional details that are not captured by this simplified model, but generate specific features in the measurements. To account for these features one has to enrich the modeling. However, this typically increases significantly the number of unknown parameters and may require fine tunning these parameters (which span an extremely large parameter space). Here, instead, we focus on the robust and rather generic features predicted by the simplified model. 

We conclude with two examples of experimental features that are not captured by the simplified model, but could be accounted for by including additional ingredients into the theoretical treatment of the problem. First, we note that the experimentally measured differential conductance is characterized by resonances in the bias voltage $V$- $BG1$ plane which breaks particle-hole symmetry (see Fig. \ref{Fig.S7}). By contrast, the calculated differential conductance is characterized by particle-hole symmetric ``parabolas''. One can introduce particle-hole asymmetry by adding dissipation, although the details of the dissipation mechanism responsible for the observed asymmetry are not clear. Alternatively, one can consider that in the high-barrier, low-tunneling regime the differential conductance is related to the local density of states (LDOS); the presence of dissipation corresponds to only consider the particle contribution to the LDOS (i.e., neglecting the hole contribution). The dependence of the LDOS (including the contribution from the particle sector) on the gate potential $\mu_{BG1}$ and the bias voltage $V$ is shown in Fig. \ref{TS6}. The key message is that observing ``stripy''  features like those in Fig. \ref{Fig.S7} is consistent with tunneling into a discrete set of states that have most of their spectral weight inside the segment of the wire above $BG1$ and couple strongly with the gate potential. The absence of the resonances below a certain value of $\mu_{BG1}$ signals that the chemical potential has reached the bottom of the subband. In the regime corresponding to $\delta\mu \sim \Delta$ (i.e., chemical potential close to the bottom of the subband) one expects the emergence of well separated Majorana bound states at finite magnetic fields. In turn, these weakly overlapping Majorana modes are responsible for (nearly) zero bias peaks in the differential conductance that are pinned near zero bias over a significant range of gate voltages (see Fig. \ref{TS3}) and Zeeman fields (Fig. \ref{TS4}).

The second example concerns the brightness of the ``first'' resonance, i.e., the stripe corresponding to the lowest values of $BG1$. In experiment, the first resonance(marked by the solid line) is much brighter than the next few ones, as shown in Fig. \ref{Fig.S7}. By contrast, in the calculations the first resonance is the weakest (see, for example, Fig. \ref{TS6}). In general, the strength of both differential conductance and LDOS is determined by the amplitude of the wave function inside the tunnel barrier region. Typically, higher energy states, which have Fourier components associated with higher values of the wave vector (i.e., loosely speaking, larger values of the Fermi momentum), penetrate deeper into the barrier region and, consequently have higher visibility. This is, indeed, the case if we assume that the step-like potential has a constant value throughout the $BG1$ region. However, the exact potential profile, which cannot be determined experimentally and is extremely difficult to calculate, may be slightly different. If, for example, we assume that the $BG1$ potential, instead of being flat, has a small dip as one approaches the tunnel barrier (see fig. \ref{TS7}), the brightness of the low-energy states changes significantly. In particular, the first resonance becomes much stronger. This is explained by the fact that in the presence of the non-homogeneous potential on bottom gates, the lowest energy state develops a local maximum near the potential dip, which results in a significant increase of its penetration into the barrier region. This example illustrates very clearly that in order to account for all the details of the experimentally observed features it is essential to incorporate information about the nonuniform background potential into the theoretical model. As a final note, we point out that because the semiconductor wire is partially covered by a superconductor the (effective) background potential will also have a nontrivial transverse profile. Moreover, this profile will be different for different cross-sections along the wire. This is expected to generate a position-dependent spin-orbit coupling strength as well as a position-dependent induced pair potential. All these effects could have an impact on the experimentally-measured quantities. Nonetheless, the picture revealed by the simple model described here, which, in essence, predicts the emergence of weakly overlapping Majorana bound states, is expected to be rather generic and robust, as long as the non-homogeneity of the system does not exceed a certain threshold.

\bigskip
\bibliographystyle{plainnat}
\bibliography{Ref_TSC.bib}

\begin{thebibliography}{10}
\expandafter\ifx\csname url\endcsname\relax
  \def\url#1{\texttt{#1}}\fi
\expandafter\ifx\csname urlprefix\endcsname\relax\def\urlprefix{URL }\fi
\providecommand{\bibinfo}[2]{#2}
\providecommand{\eprint}[2][]{\url{#2}}

\bibitem{ReadPRB2000}
\bibinfo{author}{Read, N.} \& \bibinfo{author}{Green, D.}
\newblock \bibinfo{title}{Paired states of fermions in two dimensions with
  breaking of parity and time-reversal symmetries and the fractional quantum
  hall effect}.
\newblock \emph{\bibinfo{journal}{Physical Review B}}
  \textbf{\bibinfo{volume}{61}}, \bibinfo{pages}{10267} (\bibinfo{year}{2000}).

\bibitem{KitaevAP03}
\bibinfo{author}{Kitaev, A.~Y.}
\newblock \bibinfo{title}{Fault-tolerant quantum computation by anyons}.
\newblock \emph{\bibinfo{journal}{Annals of Physics}}
  \textbf{\bibinfo{volume}{303}}, \bibinfo{pages}{2--30}
  (\bibinfo{year}{2003}).

\bibitem{NayakRMP2008}
\bibinfo{author}{Nayak, C.}, \bibinfo{author}{Simon, S.~H.},
  \bibinfo{author}{Stern, A.}, \bibinfo{author}{Freedman, M.} \&
  \bibinfo{author}{Sarma, S.~D.}
\newblock \bibinfo{title}{Non-abelian anyons and topological quantum
  computation}.
\newblock \emph{\bibinfo{journal}{Reviews of Modern Physics}}
  \textbf{\bibinfo{volume}{80}}, \bibinfo{pages}{1083} (\bibinfo{year}{2008}).

\bibitem{KitaevPU2001}
\bibinfo{author}{Kitaev, A.~Y.}
\newblock \bibinfo{title}{Unpaired majorana fermions in quantum wires}.
\newblock \emph{\bibinfo{journal}{Physics-Uspekhi}}
  \textbf{\bibinfo{volume}{44}}, \bibinfo{pages}{131} (\bibinfo{year}{2001}).

\bibitem{MourikScience2012}
\bibinfo{author}{Mourik, V.} \emph{et~al.}
\newblock \bibinfo{title}{Signatures of majorana fermions in hybrid
  superconductor-semiconductor nanowire devices}.
\newblock \emph{\bibinfo{journal}{Science}} \textbf{\bibinfo{volume}{336}},
  \bibinfo{pages}{1003--1007} (\bibinfo{year}{2012}).

\bibitem{Nadj-PergeScience2014}
\bibinfo{author}{Nadj-Perge, S.} \emph{et~al.}
\newblock \bibinfo{title}{Observation of majorana fermions in ferromagnetic
  atomic chains on a superconductor}.
\newblock \emph{\bibinfo{journal}{Science}} \textbf{\bibinfo{volume}{346}},
  \bibinfo{pages}{602--607} (\bibinfo{year}{2014}).

\bibitem{LutchynPRL2010}
\bibinfo{author}{Lutchyn, R.~M.}, \bibinfo{author}{Sau, J.~D.} \&
  \bibinfo{author}{Das~Sarma, S.}
\newblock \bibinfo{title}{Majorana fermions and a topological phase transition
  in semiconductor-superconductor heterostructures}.
\newblock \emph{\bibinfo{journal}{Physical review letters}}
  \textbf{\bibinfo{volume}{105}}, \bibinfo{pages}{077001}
  (\bibinfo{year}{2010}).

\bibitem{OregPRL2010}
\bibinfo{author}{Oreg, Y.}, \bibinfo{author}{Refael, G.} \&
  \bibinfo{author}{von Oppen, F.}
\newblock \bibinfo{title}{Helical liquids and majorana bound states in quantum
  wires}.
\newblock \emph{\bibinfo{journal}{Physical review letters}}
  \textbf{\bibinfo{volume}{105}}, \bibinfo{pages}{177002}
  (\bibinfo{year}{2010}).

\bibitem{PotterPRB2011}
\bibinfo{author}{Potter, A.~C.} \& \bibinfo{author}{Lee, P.~A.}
\newblock \bibinfo{title}{Majorana end states in multiband microstructures with
  rashba spin-orbit coupling}.
\newblock \emph{\bibinfo{journal}{Physical Review B}}
  \textbf{\bibinfo{volume}{83}}, \bibinfo{pages}{094525}
  (\bibinfo{year}{2011}).

\bibitem{StanescuPRB2011}
\bibinfo{author}{Stanescu, T.~D.}, \bibinfo{author}{Lutchyn, R.~M.} \&
  \bibinfo{author}{Sarma, S.~D.}
\newblock \bibinfo{title}{Majorana fermions in semiconductor nanowires}.
\newblock \emph{\bibinfo{journal}{Physical Review B}}
  \textbf{\bibinfo{volume}{84}}, \bibinfo{pages}{144522}
  (\bibinfo{year}{2011}).

\bibitem{MishmashPRB2016}
\bibinfo{author}{Mishmash, R.~V.}, \bibinfo{author}{Aasen, D.},
  \bibinfo{author}{Higginbotham, A.~P.} \& \bibinfo{author}{Alicea, J.}
\newblock \bibinfo{title}{Approaching a topological phase transition in
  majorana nanowires}.
\newblock \emph{\bibinfo{journal}{Physical Review B}}
  \textbf{\bibinfo{volume}{93}}, \bibinfo{pages}{245404}
  (\bibinfo{year}{2016}).

\bibitem{AliceaNatphy2011}
\bibinfo{author}{Alicea, J.}, \bibinfo{author}{Oreg, Y.},
  \bibinfo{author}{Refael, G.}, \bibinfo{author}{von Oppen, F.} \&
  \bibinfo{author}{Fisher, M.~P.}
\newblock \bibinfo{title}{Non-abelian statistics and topological quantum
  information processing in 1d wire networks}.
\newblock \emph{\bibinfo{journal}{Nature Physics}}
  \textbf{\bibinfo{volume}{7}}, \bibinfo{pages}{412--417}
  (\bibinfo{year}{2011}).

\bibitem{vanHeckNJP2012}
\bibinfo{author}{Van~Heck, B.}, \bibinfo{author}{Akhmerov, A.},
  \bibinfo{author}{Hassler, F.}, \bibinfo{author}{Burrello, M.} \&
  \bibinfo{author}{Beenakker, C.}
\newblock \bibinfo{title}{Coulomb-assisted braiding of majorana fermions in a
  josephson junction array}.
\newblock \emph{\bibinfo{journal}{New Journal of Physics}}
  \textbf{\bibinfo{volume}{14}}, \bibinfo{pages}{035019}
  (\bibinfo{year}{2012}).

\bibitem{LeeNatnano2014}
\bibinfo{author}{Lee, E.~J.} \emph{et~al.}
\newblock \bibinfo{title}{Spin-resolved andreev levels and parity crossings in
  hybrid superconductor-semiconductor nanostructures}.
\newblock \emph{\bibinfo{journal}{Nature nanotechnology}}
  \textbf{\bibinfo{volume}{9}}, \bibinfo{pages}{79--84} (\bibinfo{year}{2014}).

\bibitem{KouwenhovenRPP01}
\bibinfo{author}{Kouwenhoven, L.~P.}, \bibinfo{author}{Austing, D.~G.} \&
  \bibinfo{author}{Tarucha, S.}
\newblock \bibinfo{title}{Few-electron quantum dots}.
\newblock \emph{\bibinfo{journal}{Reports on Progress in Physics}}
  \textbf{\bibinfo{volume}{64}}, \bibinfo{pages}{701} (\bibinfo{year}{2001}).

\bibitem{AlbrechtNature2016}
\bibinfo{author}{Albrecht, S.~M.} \emph{et~al.}
\newblock \bibinfo{title}{Exponential protection of zero modes in majorana
  islands}.
\newblock \emph{\bibinfo{journal}{Nature}} \textbf{\bibinfo{volume}{531}},
  \bibinfo{pages}{206--209} (\bibinfo{year}{2016}).

\bibitem{PradaPRB2012}
\bibinfo{author}{Prada, E.}, \bibinfo{author}{San-Jose, P.} \&
  \bibinfo{author}{Aguado, R.}
\newblock \bibinfo{title}{Transport spectroscopy of n s nanowire junctions with
  majorana fermions}.
\newblock \emph{\bibinfo{journal}{Physical Review B}}
  \textbf{\bibinfo{volume}{86}}, \bibinfo{pages}{180503}
  (\bibinfo{year}{2012}).

\bibitem{DasSarmaPRB2012}
\bibinfo{author}{Sarma, S.~D.}, \bibinfo{author}{Sau, J.~D.} \&
  \bibinfo{author}{Stanescu, T.~D.}
\newblock \bibinfo{title}{Splitting of the zero-bias conductance peak as
  smoking gun evidence for the existence of the majorana mode in a
  superconductor-semiconductor nanowire}.
\newblock \emph{\bibinfo{journal}{Physical Review B}}
  \textbf{\bibinfo{volume}{86}}, \bibinfo{pages}{220506}
  (\bibinfo{year}{2012}).

\bibitem{StanescuPRB2013}
\bibinfo{author}{Stanescu, T.~D.}, \bibinfo{author}{Lutchyn, R.~M.} \&
  \bibinfo{author}{Sarma, S.~D.}
\newblock \bibinfo{title}{Dimensional crossover in spin-orbit-coupled
  semiconductor nanowires with induced superconducting pairing}.
\newblock \emph{\bibinfo{journal}{Physical Review B}}
  \textbf{\bibinfo{volume}{87}}, \bibinfo{pages}{094518}
  (\bibinfo{year}{2013}).

\bibitem{RainisPRB2013}
\bibinfo{author}{Rainis, D.}, \bibinfo{author}{Trifunovic, L.},
  \bibinfo{author}{Klinovaja, J.} \& \bibinfo{author}{Loss, D.}
\newblock \bibinfo{title}{Towards a realistic transport modeling in a
  superconducting nanowire with majorana fermions}.
\newblock \emph{\bibinfo{journal}{Physical Review B}}
  \textbf{\bibinfo{volume}{87}}, \bibinfo{pages}{024515}
  (\bibinfo{year}{2013}).

\bibitem{Nadj-PergePRL12}
\bibinfo{author}{Nadj-Perge, S.} \emph{et~al.}
\newblock \bibinfo{title}{Spectroscopy of spin-orbit quantum bits in indium
  antimonide nanowires}.
\newblock \emph{\bibinfo{journal}{Phys. Rev. Lett.}}
  \textbf{\bibinfo{volume}{108}}, \bibinfo{pages}{166801}
  (\bibinfo{year}{2012}).

\bibitem{linPRB12}
\bibinfo{author}{Lin, C.-H.}, \bibinfo{author}{Sau, J.~D.} \&
  \bibinfo{author}{Das~Sarma, S.}
\newblock \bibinfo{title}{Zero-bias conductance peak in majorana wires made of
  semiconductor/superconductor hybrid structures}.
\newblock \emph{\bibinfo{journal}{Phys. Rev. B}} \textbf{\bibinfo{volume}{86}},
  \bibinfo{pages}{224511} (\bibinfo{year}{2012}).

\bibitem{StoudenmirePRB2011}
\bibinfo{author}{Stoudenmire, E.}, \bibinfo{author}{Alicea, J.},
  \bibinfo{author}{Starykh, O.~A.} \& \bibinfo{author}{Fisher, M.~P.}
\newblock \bibinfo{title}{Interaction effects in topological superconducting
  wires supporting majorana fermions}.
\newblock \emph{\bibinfo{journal}{Physical Review B}}
  \textbf{\bibinfo{volume}{84}}, \bibinfo{pages}{014503}
  (\bibinfo{year}{2011}).

\bibitem{adagideliprb14}
\bibinfo{author}{Adagideli, i. d.~I.}, \bibinfo{author}{Wimmer, M.} \&
  \bibinfo{author}{Teker, A.}
\newblock \bibinfo{title}{Effects of electron scattering on the topological
  properties of nanowires: Majorana fermions from disorder and superlattices}.
\newblock \emph{\bibinfo{journal}{Phys. Rev. B}} \textbf{\bibinfo{volume}{89}},
  \bibinfo{pages}{144506} (\bibinfo{year}{2014}).

\bibitem{NijholtPRB2016}
\bibinfo{author}{Nijholt, B.} \& \bibinfo{author}{Akhmerov, A.~R.}
\newblock \bibinfo{title}{Orbital effect of magnetic field on the majorana
  phase diagram}.
\newblock \emph{\bibinfo{journal}{Preprint arXiv: 1509.02675}}
  (\bibinfo{year}{2015}).

\bibitem{vuiknjp2016}
\bibinfo{author}{Vuik, A.}, \bibinfo{author}{Eeltink, D.},
  \bibinfo{author}{Akhmerov, A.} \& \bibinfo{author}{Wimmer, M.}
\newblock \bibinfo{title}{Effects of the electrostatic environment on the
  majorana nanowire devices}.
\newblock \emph{\bibinfo{journal}{New Journal of Physics}}
  \textbf{\bibinfo{volume}{18}}, \bibinfo{pages}{033013}
  (\bibinfo{year}{2016}).

\bibitem{ZhangArxiv2016}
\bibinfo{author}{Zhang, H.} \emph{et~al.}
\newblock \bibinfo{title}{Ballistic majorana nanowire devices}.
\newblock \emph{\bibinfo{journal}{Preprint arXiv: 1603.04069}}
  (\bibinfo{year}{2016}).

\bibitem{vanweperennanolett13}
\bibinfo{author}{van Weperen, I.}, \bibinfo{author}{Plissard, S.~R.},
  \bibinfo{author}{Bakkers, E. P. A.~M.}, \bibinfo{author}{Frolov, S.~M.} \&
  \bibinfo{author}{Kouwenhoven, L.~P.}
\newblock \bibinfo{title}{Quantized conductance in an insb nanowire}.
\newblock \emph{\bibinfo{journal}{Nano Letters}} \textbf{\bibinfo{volume}{13}},
  \bibinfo{pages}{387--391} (\bibinfo{year}{2013}).

\bibitem{Kammhubernanolett16}
\bibinfo{author}{Kammhuber, J.} \emph{et~al.}
\newblock \bibinfo{title}{Conductance quantization at zero magnetic field in
  insb nanowires}.
\newblock \emph{\bibinfo{journal}{Nano Letters}} \textbf{\bibinfo{volume}{16}},
  \bibinfo{pages}{3482--3486} (\bibinfo{year}{2016}).

\bibitem{Vanweperenprb15}
\bibinfo{author}{van Weperen, I.} \emph{et~al.}
\newblock \bibinfo{title}{Spin-orbit interaction in insb nanowires}.
\newblock \emph{\bibinfo{journal}{Phys. Rev. B}} \textbf{\bibinfo{volume}{91}},
  \bibinfo{pages}{201413} (\bibinfo{year}{2015}).

\bibitem{BlonderPhysRevB1982}
\bibinfo{author}{Blonder, G.}, \bibinfo{author}{Tinkham, M.} \&
  \bibinfo{author}{Klapwijk, T.}
\newblock \bibinfo{title}{Transition from metallic to tunneling regimes in
  superconducting microconstrictions: Excess current, charge imbalance, and
  supercurrent conversion}.
\newblock \emph{\bibinfo{journal}{Physical Review B}}
  \textbf{\bibinfo{volume}{25}}, \bibinfo{pages}{4515} (\bibinfo{year}{1982}).

\end{thebibliography}

\newpage

{\centering \section{Supplementary Figures}
}

\renewcommand\thefigure{S\arabic{figure}}
\begin{figure*}[h!]
\centering
  \includegraphics[width=0.9\textwidth]{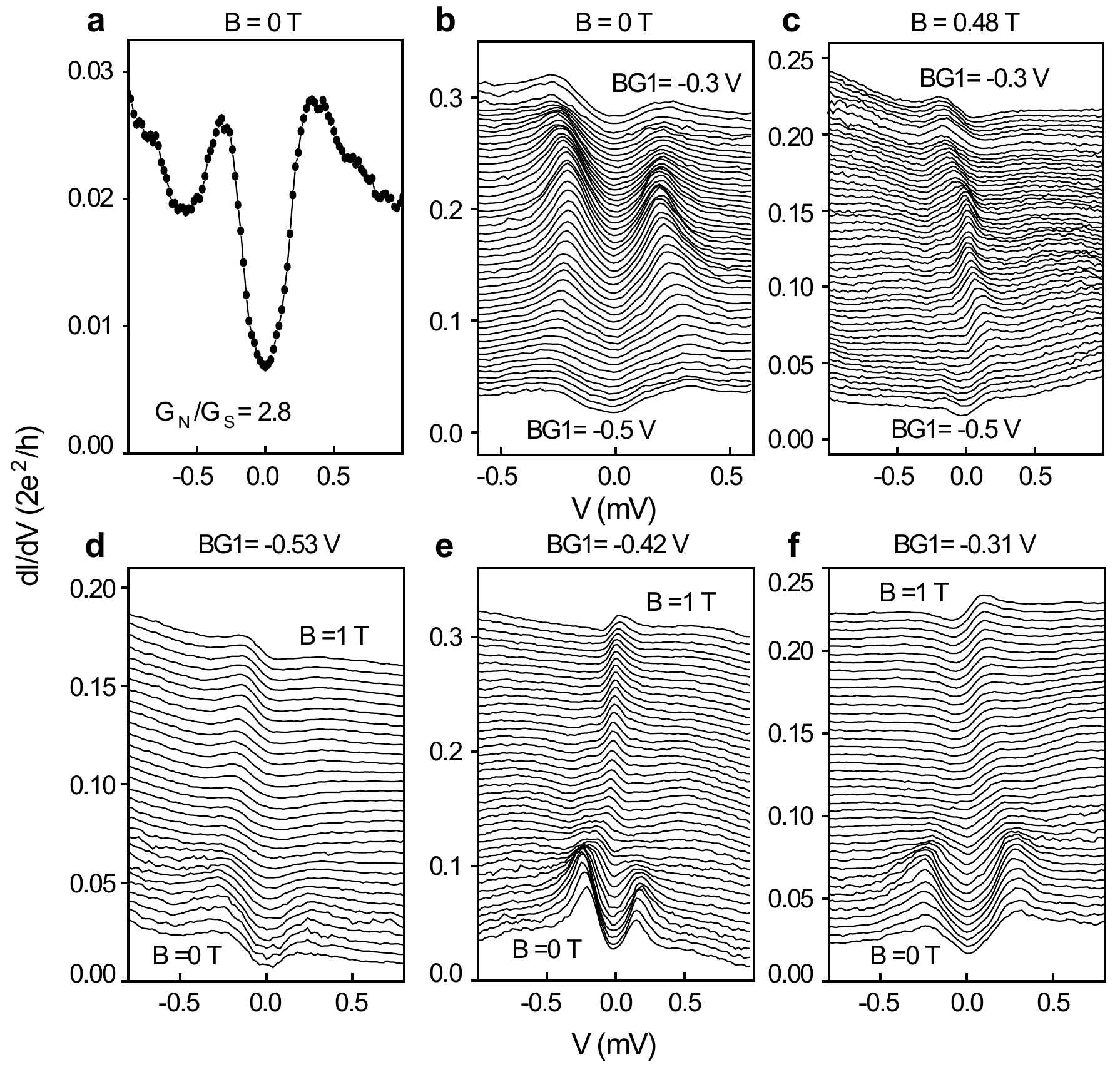}
\caption{\textbf{Linecuts from Fig. 1 and Fig. 2 in the main text.} $\bf{a}$, a linecut of bias scan at $FG1=0.14~$V and $B=0~$T. It shows a clear induced gap with gap edge at $V =~0.25~$mV. The ratio of conductance outside and inside the gap is $2.8$. $\bf{b,c}$, linecuts correspond to Fig. 2a and Fig. 2f, respectively. $\bf{d-f}$, line cuts correspond to Figs. 1c, 1d and 1e, respectively. Vertical offset is $0.005$(in the unit of $2e^{2}/h$) in Figs. \ref{Fig.S1}b-f.}
 \label{Fig.S1}
\end{figure*}


\renewcommand\thefigure{S\arabic{figure}}
\begin{figure*}[h!]
\centering
  \includegraphics[width=0.9\textwidth]{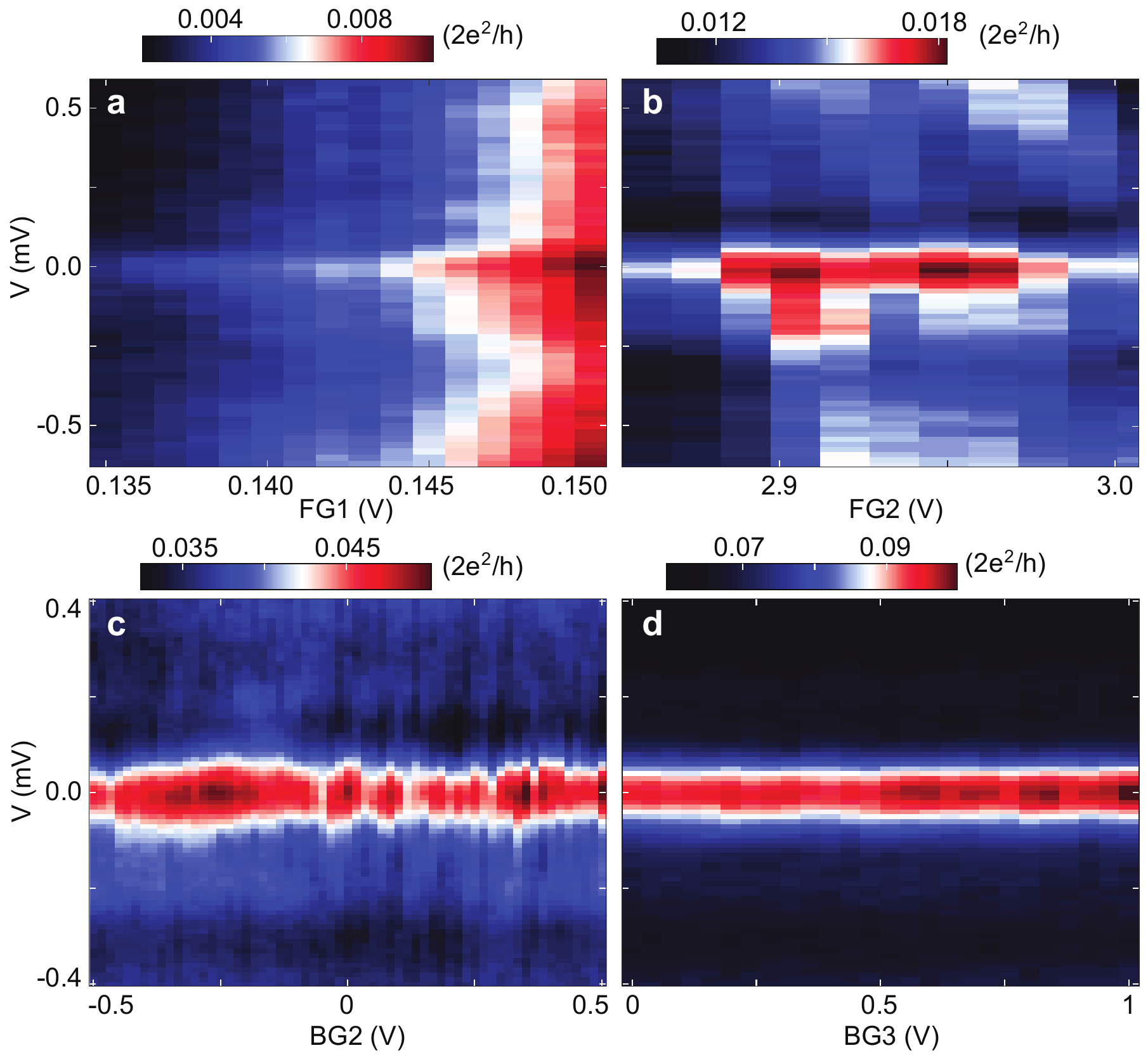}
\caption{\textbf{Gates dependence of ZBP.} $\bf{a-d}$, at fixed magnetic field $B=0.5~$T and $BG1=-0.42~$V, conductance maps are plotted in bias vs. barrier gate $FG1$, gate $FG2$ at the normal side as well as big gates $BG2$ and $BG3$ under the superconductor.}
 \label{Fig.S2}
\end{figure*}

\renewcommand\thefigure{S\arabic{figure}}
\begin{figure*}[h!]
\centering
  \includegraphics[width=0.9\textwidth]{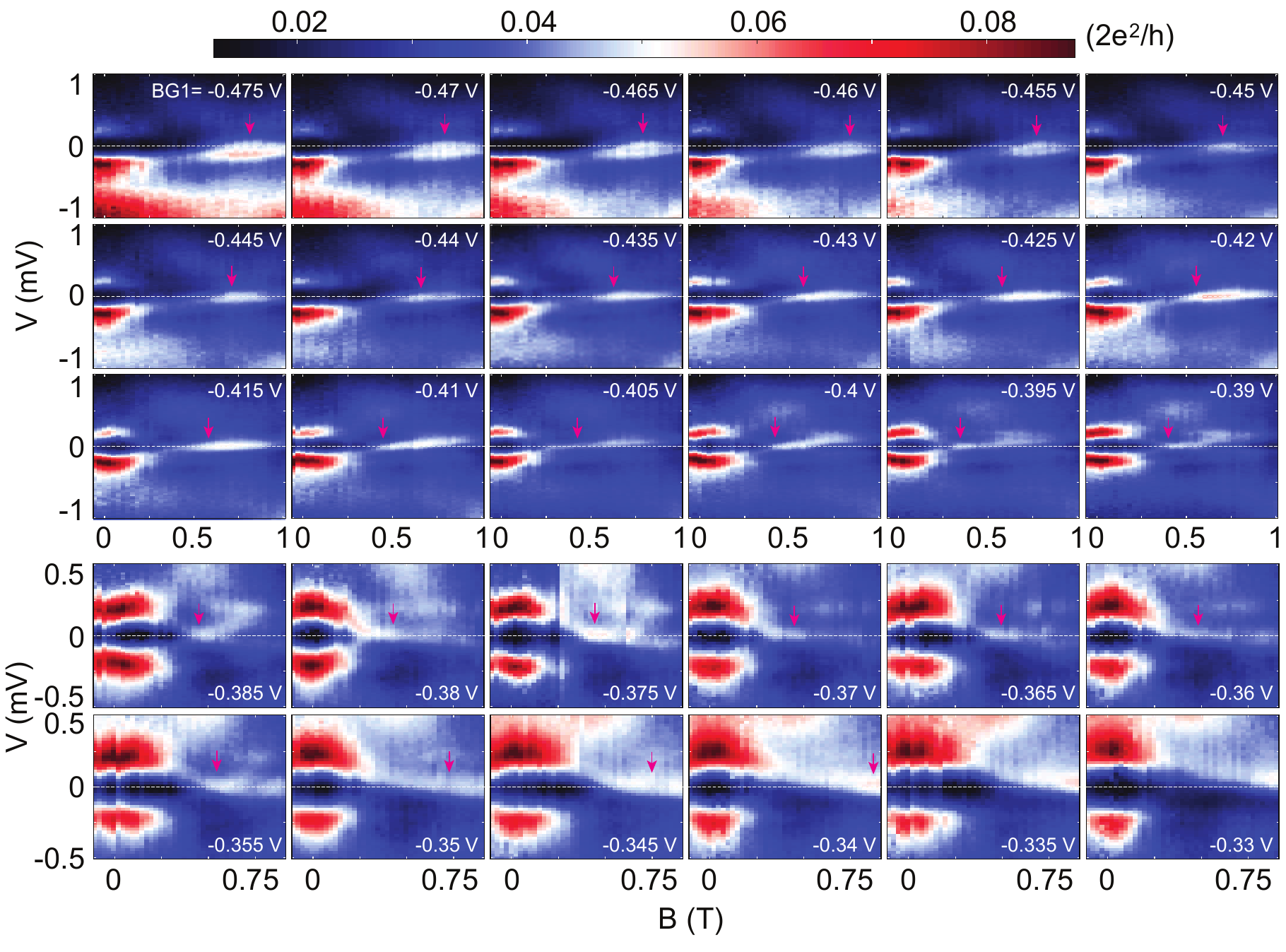}
\caption{\textbf{Zero-bias peak evolution with BG1.} Conductance maps in bias voltage $V$ vs. $B$ at different BG1 indicated in the right corner of each panel. Note that, due to a charge jump, all the gate voltages of BG1 have been shifted by $+0.02~$V. The dashed lines mark zero bias voltage line. Arrows mark the ZBP onset fields plotted in Fig. 3. The onset fields are picked by judging from line cuts of bias scans where the conductance peaks first hit zero bias voltage. 
 \label{Fig.S3}}
\end{figure*}

\renewcommand\thefigure{S\arabic{figure}}
\begin{figure*}[h!]
\centering
  \includegraphics[width=0.9\textwidth]{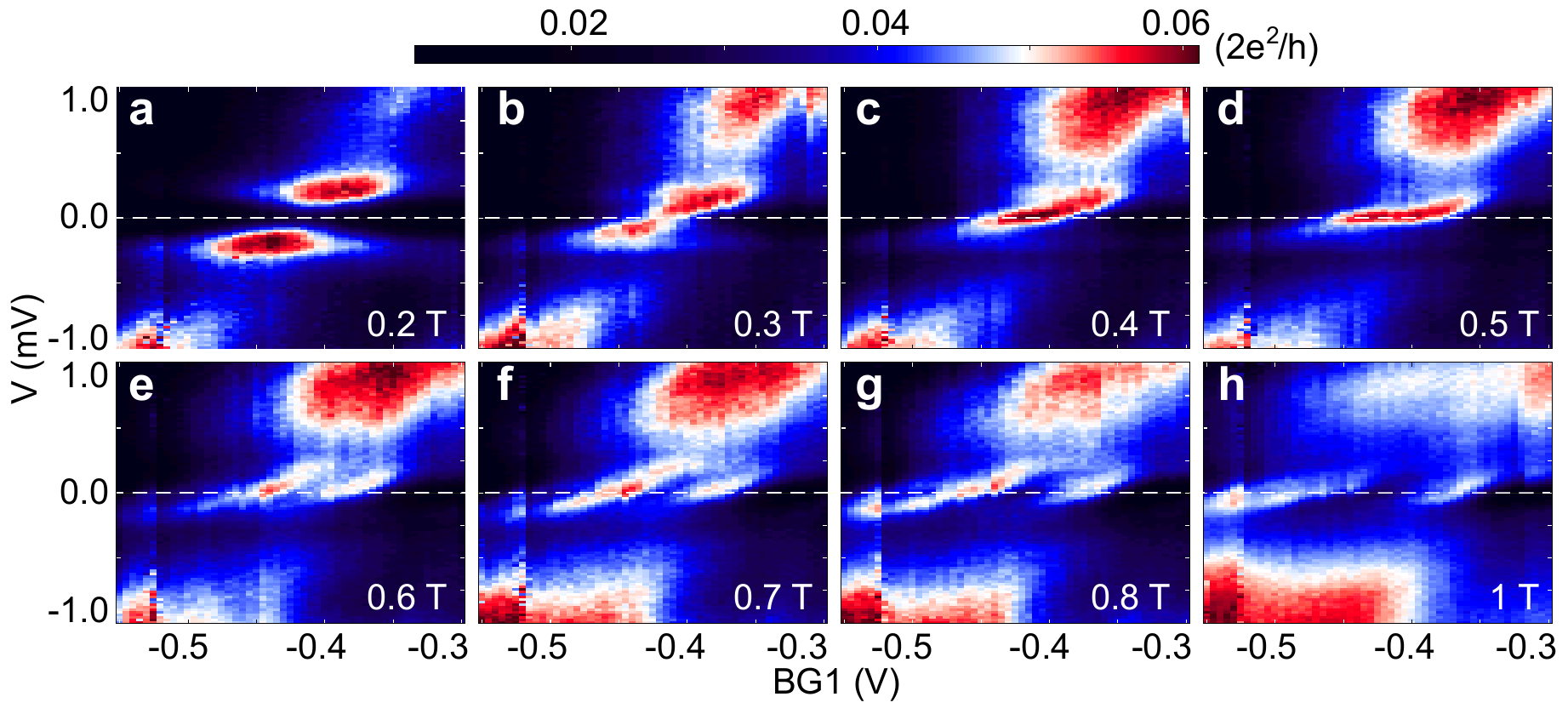}
\caption{\textbf{ZBP evolution at a large range of magnetic field.} $\bf{a-h}$, conductance maps in bias vs. BG1 at different magnetic fields indicated in lower right corners of each panel, from $0.2~$T to $1~$T. The dashed lines mark zero bias voltage line.
 \label{Fig.S4}}
\end{figure*}

\begin{figure*}[h!]
\centering
  \includegraphics[width=0.9\textwidth]{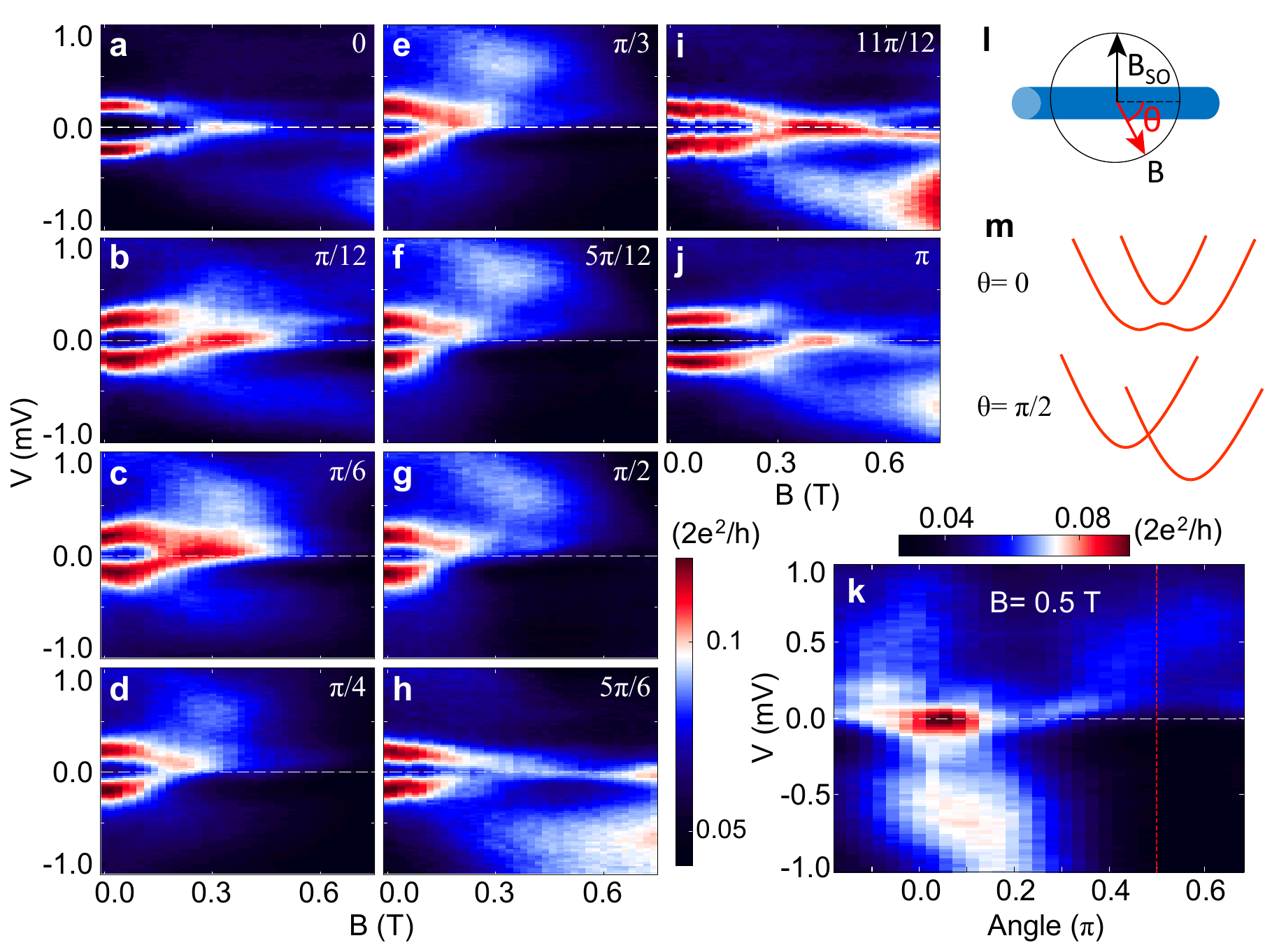}
\caption{\textbf{Magnetic field orientation dependence of ZBP.} $\bf{a-j}$, conductance maps in bias vs. magnetic field at different angles indicated in the upper right corner of each panel, from $0$ to $\pi$. $\bf{k}$, conductance maps in bias vs. field angle at a fixed field $B=0.5~$T, the vertical dashed line marks the angle at $\pi/2$. $\bf{l}$, Schematics of magnetic field direction. The angle is defined with respect to nanowire main axis. The dashed lines mark zero bias voltage line. \label{Fig.S5} $\bf{m}$, Schematics of the band structure in magnetic field at an angle of 0 and $\pi/2$.
 \label{Fig.S5}}
\end{figure*}


\begin{figure*}[h!]
\centering
  \includegraphics[width=0.9\textwidth]{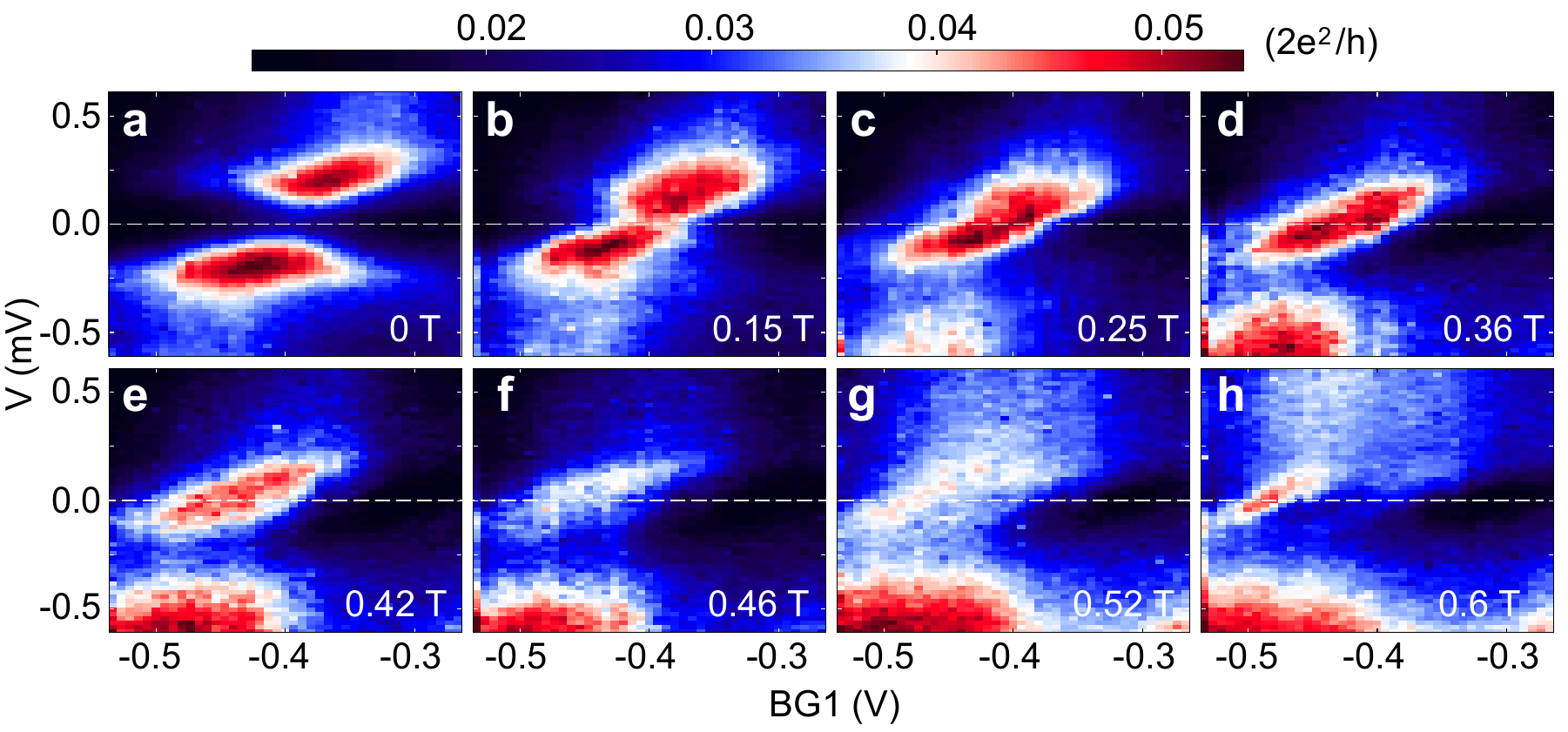}
\caption{\textbf{ZBP evolution with BG1 at an angle of $\pi/2$.} $\bf{a-h}$, conductance maps in bias vs. BG1 at different magnetic fields indicated in the lower right corner of each panel. The dashed lines mark zero bias voltage line.
 \label{Fig.S6}}
\end{figure*}


\begin{figure*}[h!]
\centering
  \includegraphics[width=0.9\textwidth]{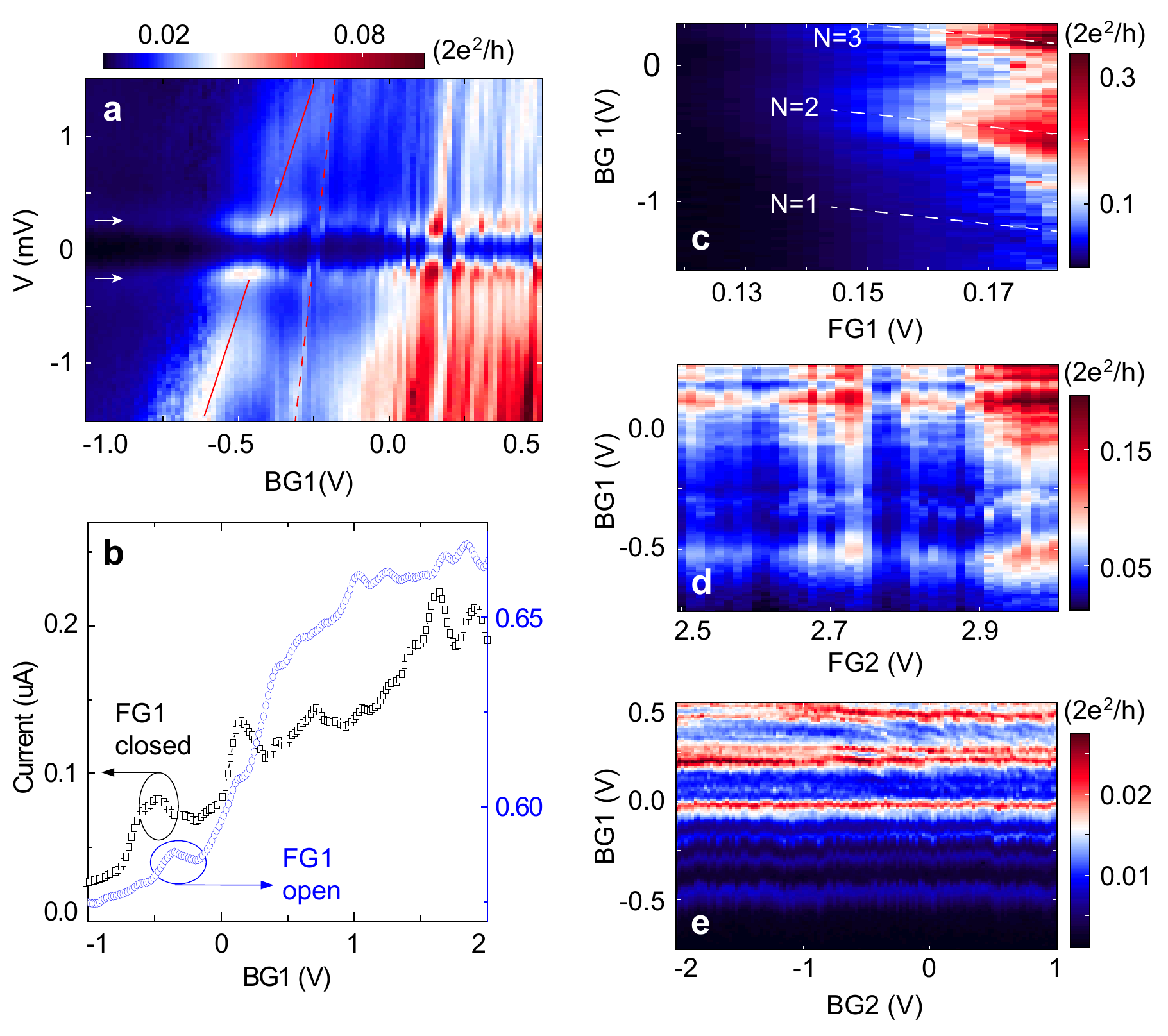}
\caption{\textbf{Expanded scan of BG1, and gates dependence of resonances.} $\bf{a}$, a conductance map in bias vs. BG1 in expanded range. The first resonance is marked by a solid line and the second one is marked by a dashed line. Dispersion of the first(solid line) and second(dashed line) resonances are $10~$meV/V and $25~$meV/V, respectively. $\bf{b}$, DC current in gate traces of BG1 with FG1 set to be closed (square) and open(circle), respectively. The voltage bias are $2~$mV and $10~$mV, respectively. $\bf{c-d}$, Dependence of the resonances at zero bias on barrier gate $FG1$, gate $FG2$ at the normal contact side, and $BG2$ under the superconductor. Dashed lines labled by $N=1$, $N=2$ and $N=3$ in $\bf{c}$ correspond to bottom of the first, second and third subband, respectively. All the scans are taken at zero magnetic field.
 \label{Fig.S7}}
\end{figure*}


\begin{figure*}[h!]
\centering
  \includegraphics[width=0.9\textwidth]{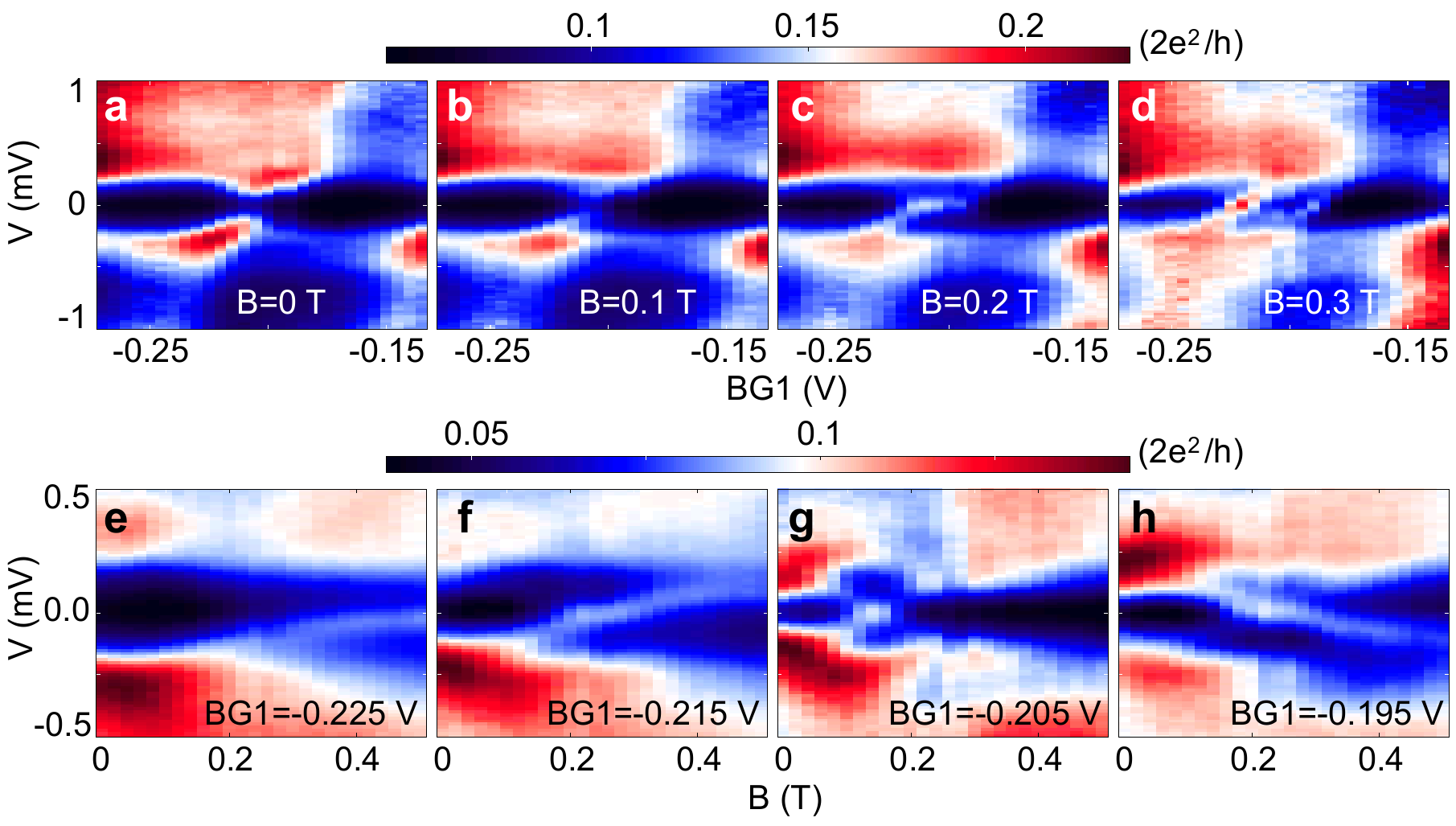}
\caption{\textbf{Gate and field dependence of the second resonance.} The second resonance is shown in Fig. \ref{Fig.S7}, marked by the dashed line. $\bf{a-d}$, conductance maps in bias vs. $BG1$ at different magnetic fields. $\bf{e-h}$, conductance maps in bias vs. field at different $BG1$. 
 \label{Fig.S8}}
\end{figure*}

\begin{figure*}[h!]
\centering
  \includegraphics[width=0.9\textwidth]{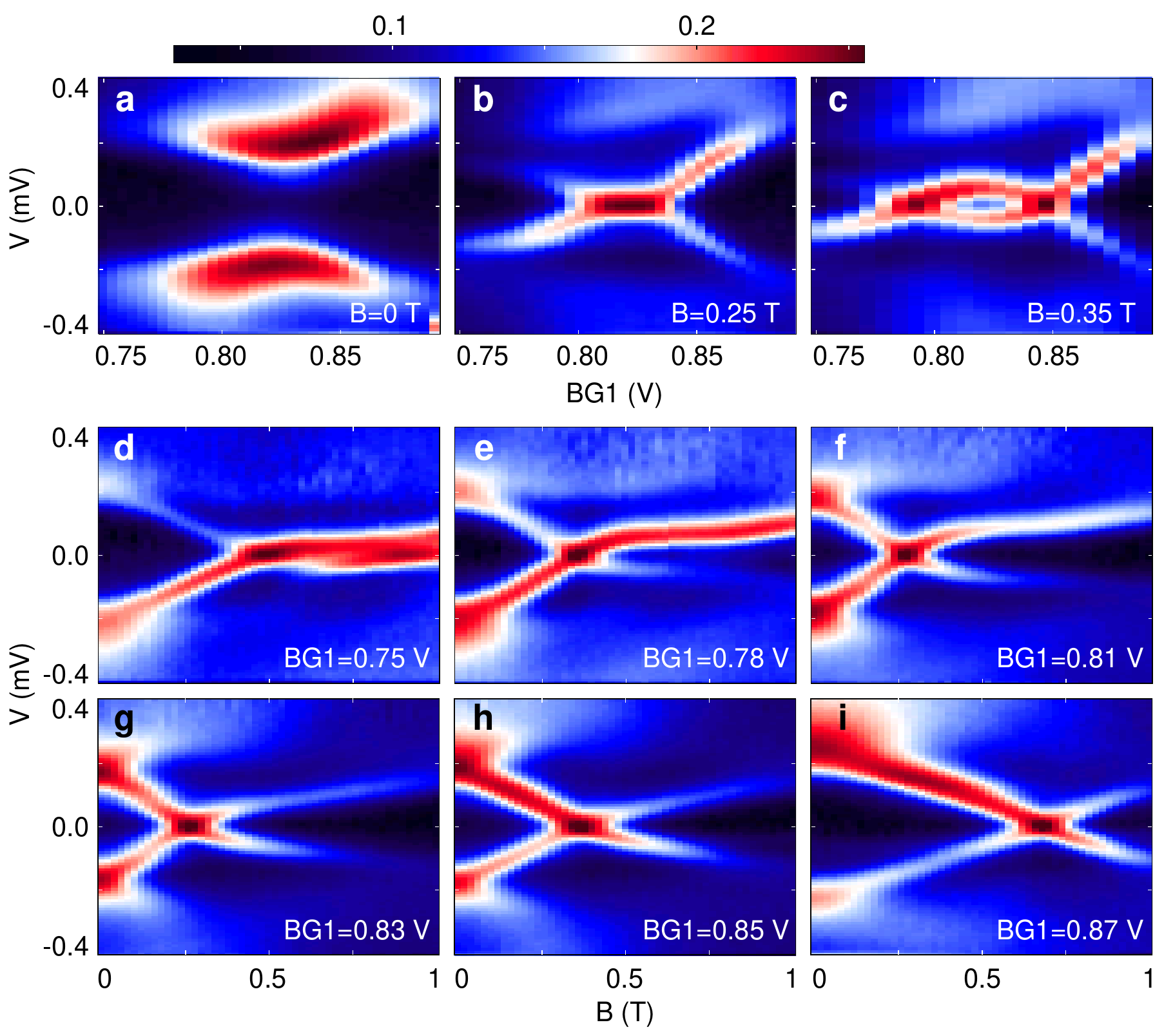}
\caption{\textbf{Trivial ABS in another device.} $\bf{a-c}$, conductance maps in bias vs. $BG1$ at three different fields indicated in the lower right corner of each panel. $\bf{d-i}$, conductance maps in bias vs. field at different $BG1$ indicated in the lower right corner of each panel.
 \label{Fig.S9}}
\end{figure*}

\begin{figure*}[h!]
\centering
  \includegraphics[width=0.9\textwidth]{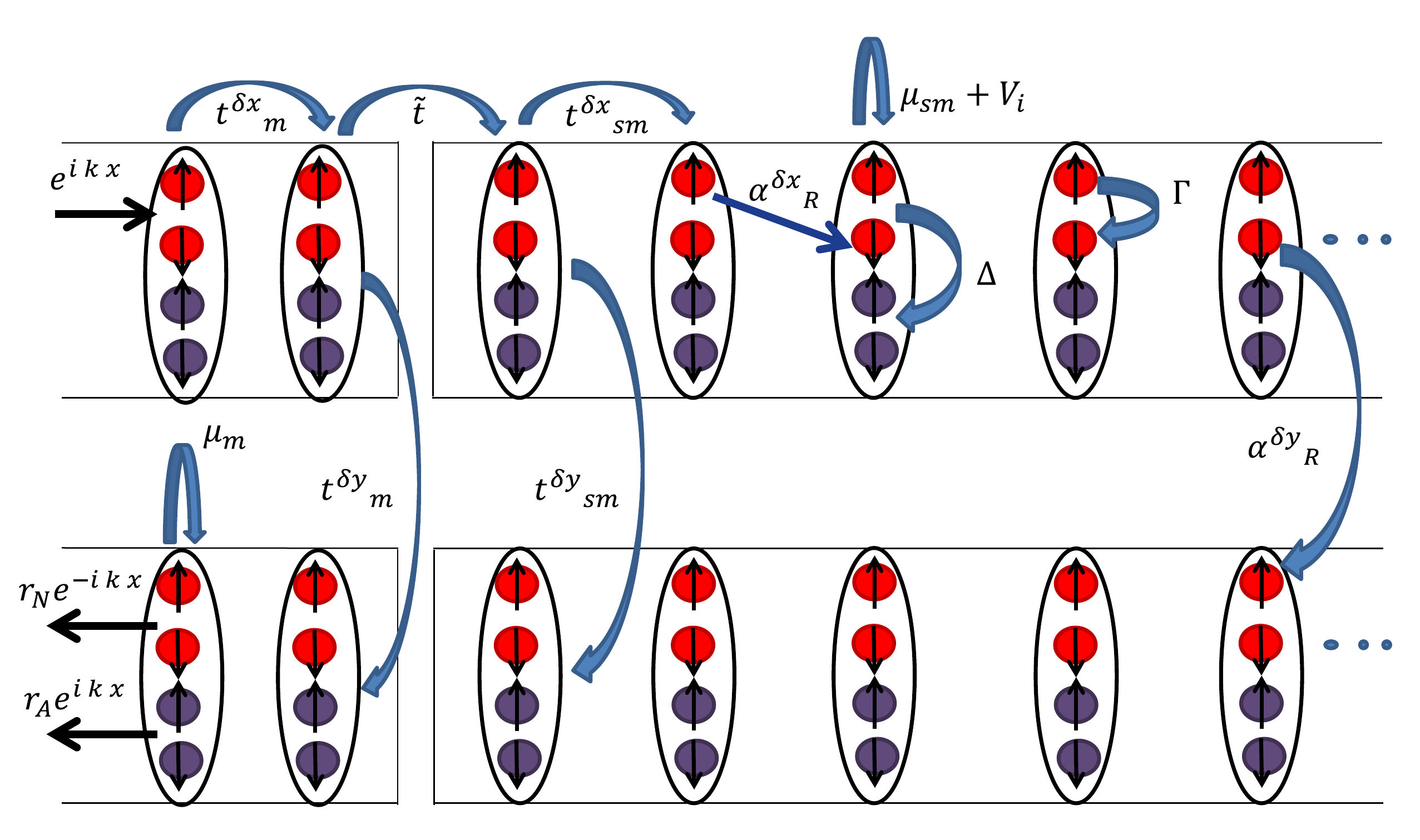}
\caption{\textbf{Schematic representation of the tight-binding Hamiltonian} $H=H_M+H_{SM}+H_{M-SM}$ given by Eqs. (\ref{EqThS1}-\ref{EqThS3}), which describes the metal-proximitized semiconductor structure. Two parallel chains ($N_y=2$) are explicitly shown and for each site (represented by an ellipse) the particle-hole and spin sectors are represented by red/blue circles and up/down arrows, respectively. The hopping parameters ($t_{sm}^\delta$, $t_{m}^\delta$, and  $\tilde{t}$), on-site and chemical potentials ($V_i$, $\mu_m$, and $\mu_{sm}$), Zeeman field ($\Gamma$), Rashba spin-orbit coupling ($\alpha^{\delta x}_R$ and $\alpha^{\delta y}_R$), and pair potential ($\Delta$) are represented by arrows that couple different chains, sites, and/or sectors.}
\label{TS1}
\end{figure*}

\begin{figure*}[h!]
\centering
  \includegraphics[width=0.9\textwidth]{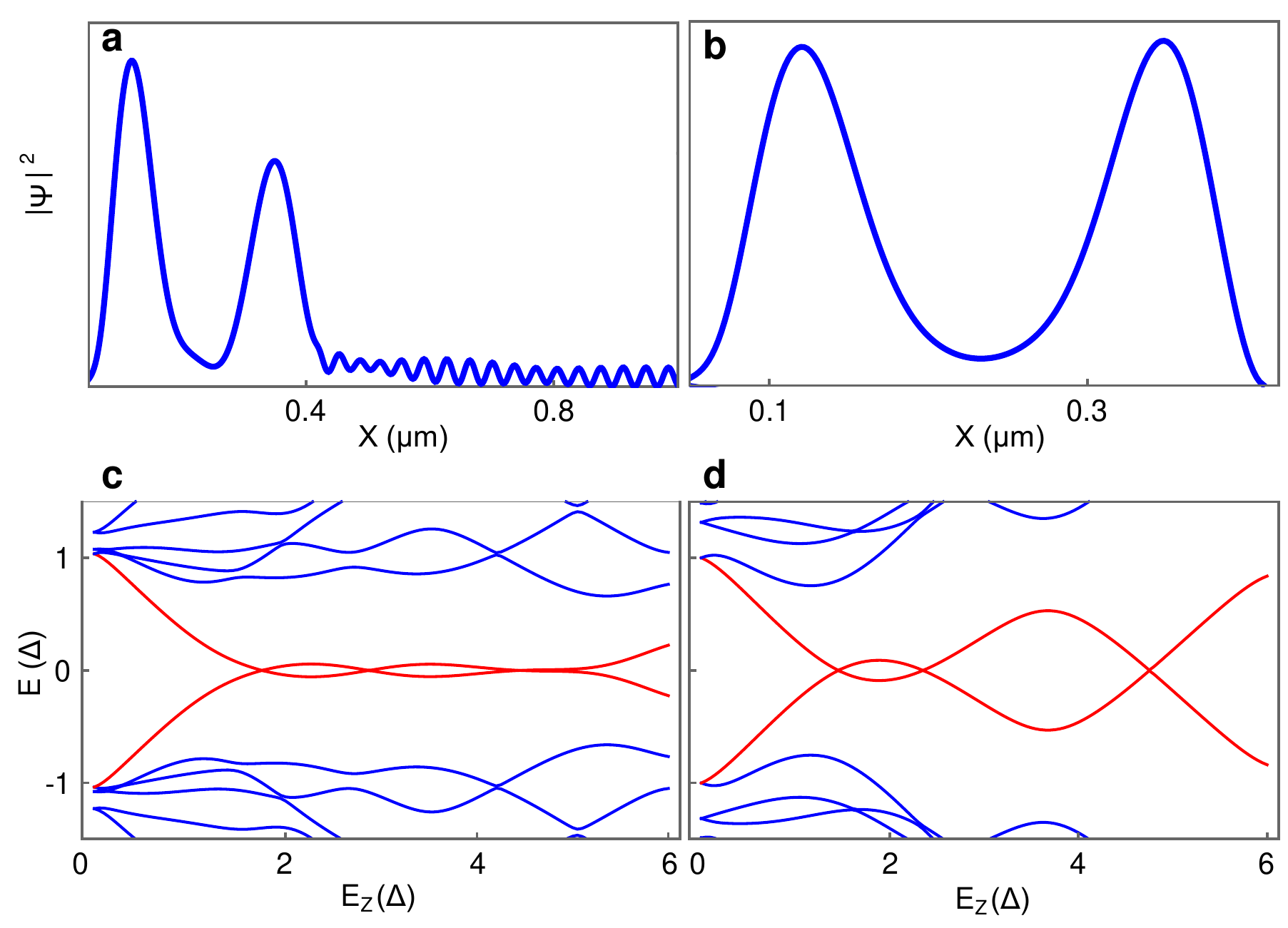}
\caption{{\em Top}: Comparison between the spatial dependence of the amplitudes of the lowest-energy states corresponding to a long wire with a step-like potential ({\bf a}) and a short wire of length equal to the $BG1$ region ({\bf b}). {\em Bottom}: Zeeman field dependence of the low-energy spectrum for a long wire with step-like potential ({\bf c}) and a short wire corresponding to the $BG1$ segment ({\bf d}). The amplitude of the zero-energy splitting oscillations show a qualitatively different dependence on the Zeeman field.}
\vspace{-2mm}
\label{TS2}
\end{figure*}

\begin{figure*}[h!]
\centering
  \includegraphics[width=0.9\textwidth]{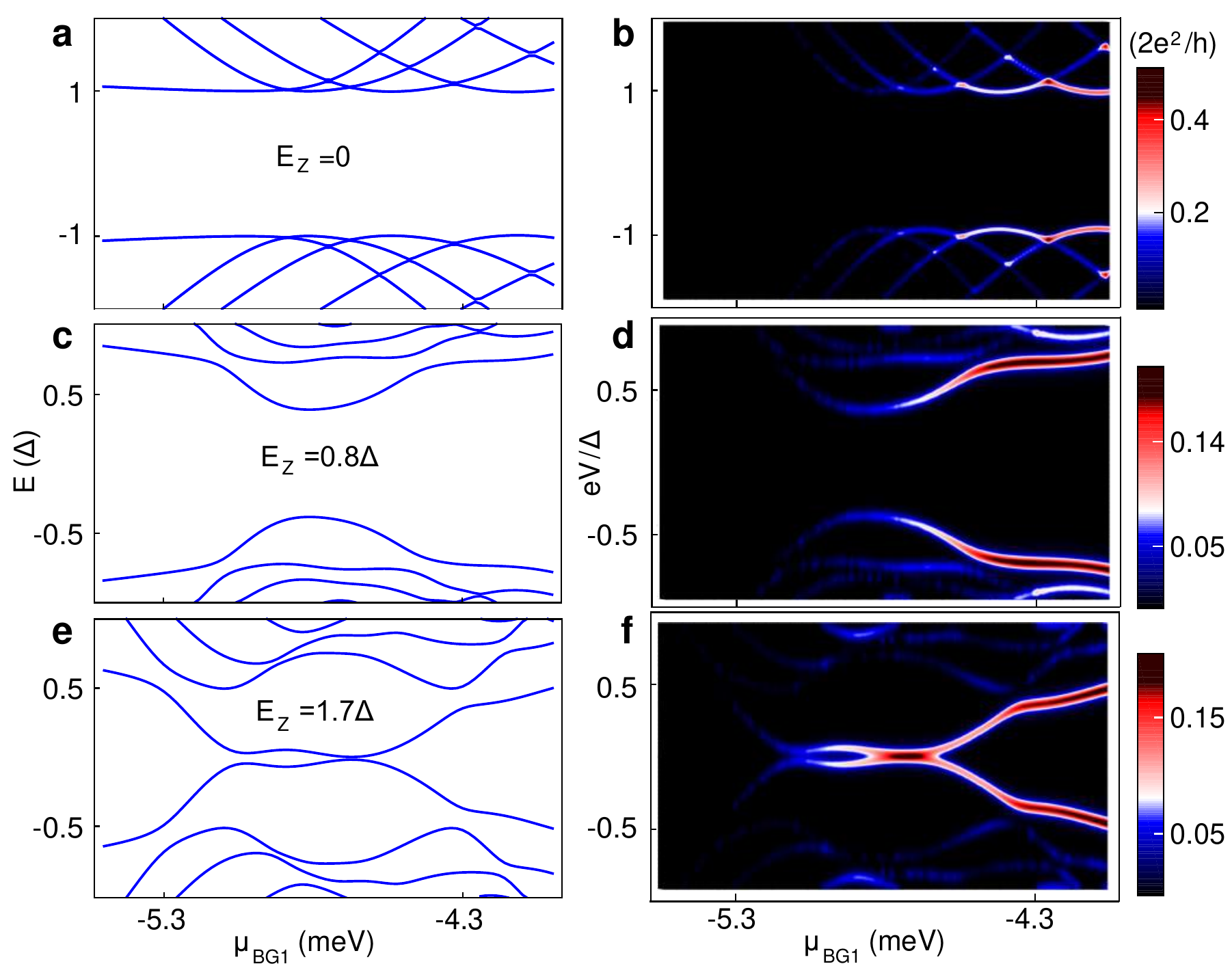}
\caption{{\em Left}: Low-energy spectrum as a function of $\mu_{BG1}$ for three different values of the Zeeman field ($E_Z=0.0~\Delta$({\bf a}), $E_Z=0.8~\Delta$({\bf c}), and $E_Z=1.7~\Delta$({\bf e})). {\em Right}:({\bf b, d, f}), Conductance maps of the differential conductance as functions of bias voltage $V$ and $\mu_{BG1}$ for the same Zeeman field as the left panels. A small thermal broadening of $0.02~$meV is used here, to emphasize the close correspondence with the corresponding spectra.}
\vspace{-2mm}
\label{TS3}
\end{figure*}


\begin{figure*}[h!]
\centering
  \includegraphics[width=0.9\textwidth]{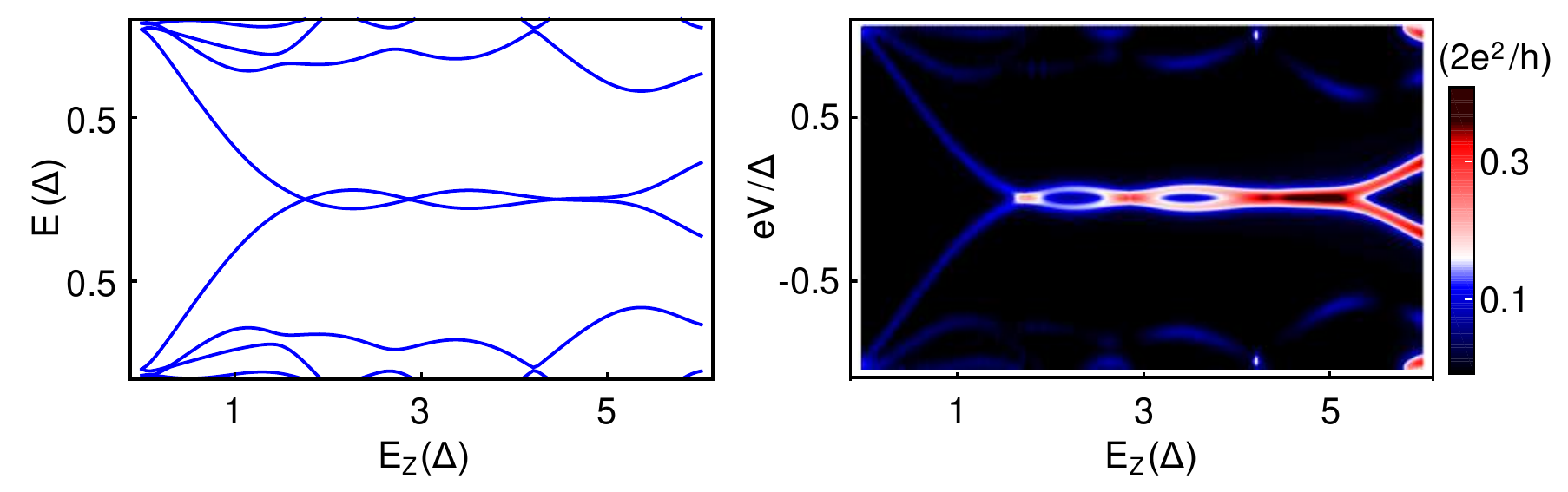}
\caption{{\bf a}, Dependence of the low-energy spectrum on the Zeeman field for a system with a step-like potential with $\mu_{BG1} =-4.7~$meV. {\bf b}, Differential conductance at $\mu_{BG1}=-4.7~$meV as functions of magnetic field and bias potential. Note that for $E_{Z}>1.7\Delta$ the system hosts nearly-zero energy modes. Also note that for $E_{Z}<5~\Delta$ the amplitude of the zero-energy splitting oscillations does not increase with the applied magnetic field.}
\vspace{-2mm}
\label{TS4}
\end{figure*}

\begin{figure*}[h!]
\centering
  \includegraphics[width=0.9\textwidth]{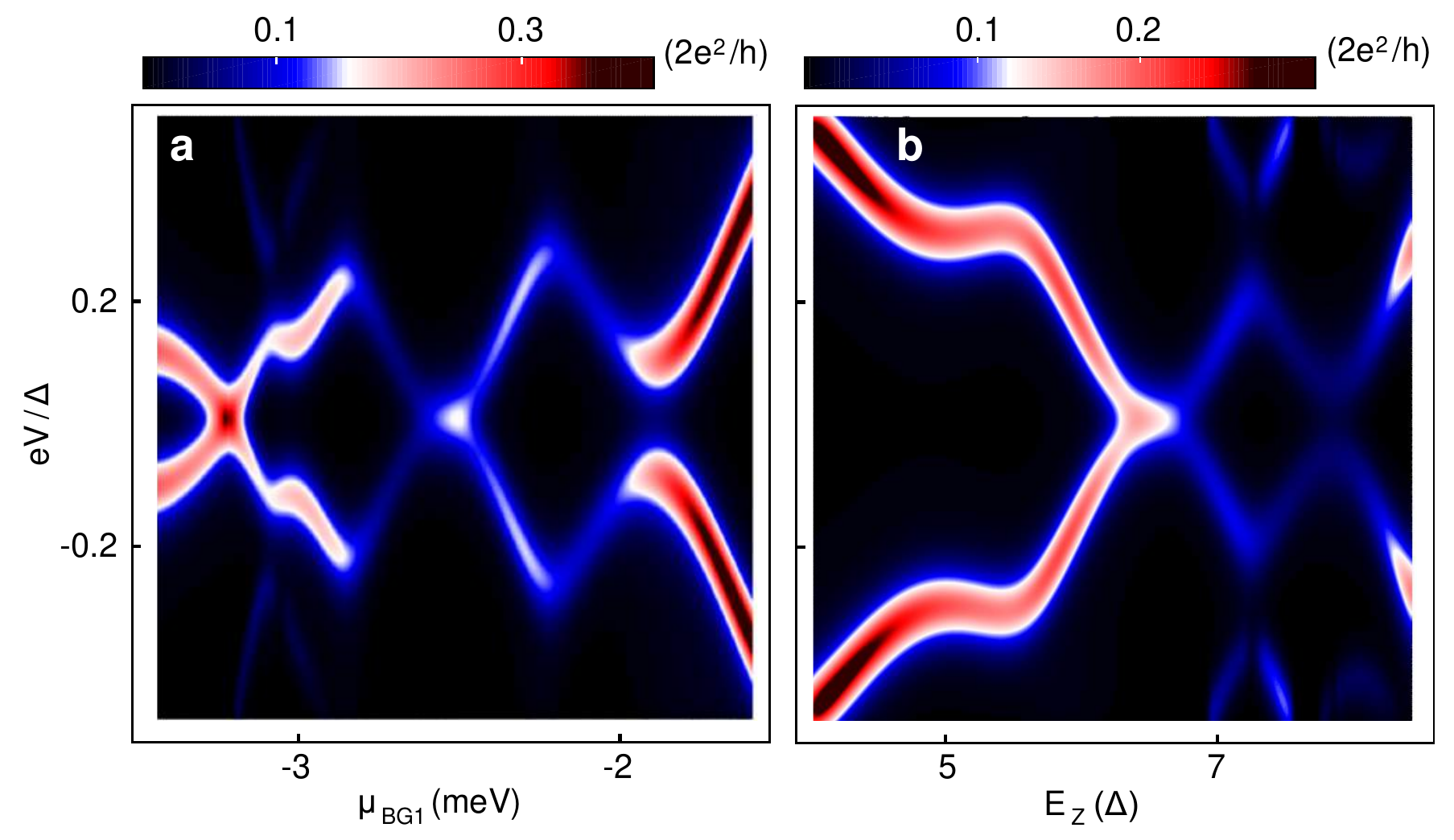}
\caption{Differential conductance at high values of $\mu_{BG1}$ corresponding to the chemical potential well above the bottom of the subband, $\delta \mu \gg \Delta$, where $\delta \mu$ is the chemical potential relative to the bottom of the band. {\bf a}, Dependence on the chemical potential at a Zeeman field of $E_Z=6.2~\Delta$. {\bf b}, Dependence on the Zeeman energy at chemical potential $\mu_{BG1}=-2.5~$meV.}
\vspace{-2mm}
\label{TS5}
\end{figure*}

\begin{figure*}[h!]
\centering
  \includegraphics[width=0.9\textwidth]{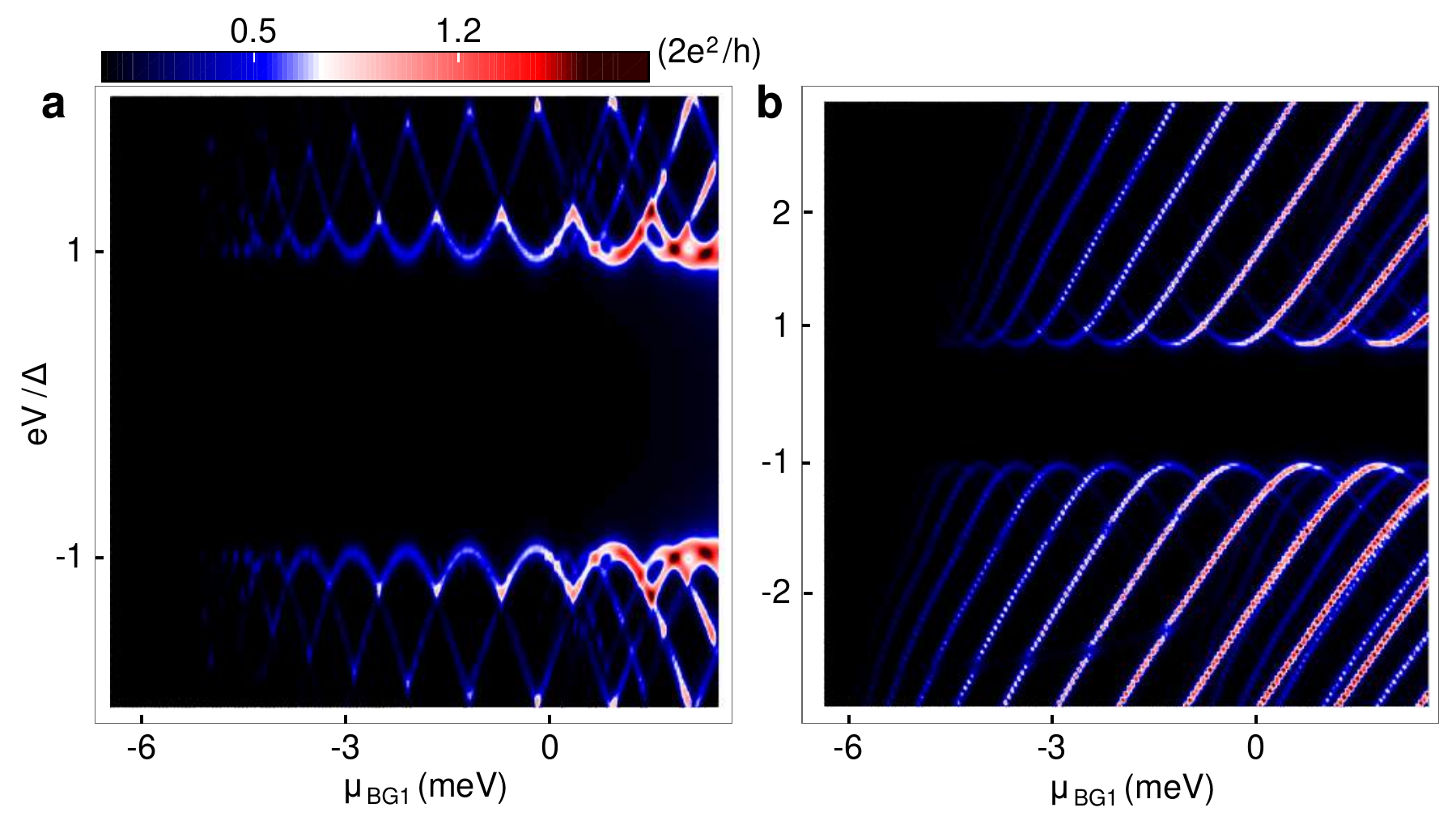}
\caption{{\bf a}, Conductance map at zero magnetic field. {\bf b}, color map of local density of particle states at zero magnetic field.}
\vspace{-2mm}
\label{TS6}
\end{figure*}

\begin{figure*}[h!]
\centering
  \includegraphics[width=0.9\textwidth]{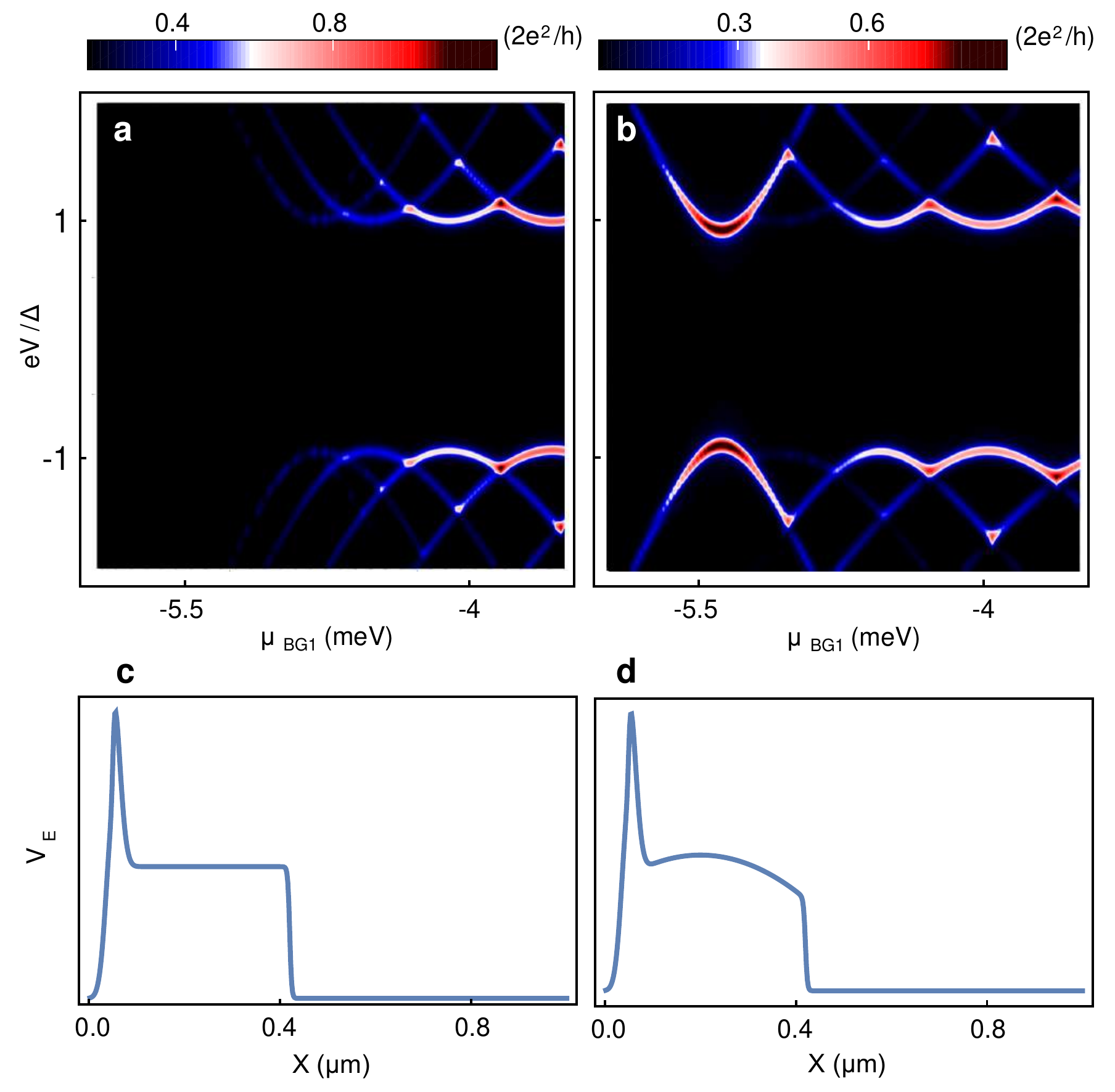}
\caption{{\bf a}, Conductance map of bias versus $\mu_{BG1}$ for a homogeneous potential. {\bf b}, differential conductance color map of bias versus $BG1$ for a non-homogeneous potential. {\bf c}, potential profile $V_{E}$ 
along the nanowire (also shown in Fig. 4a) used to generate the map in {\bf a}. {\bf d} potential profile $V_{E}$ along the nanowire used to generate the map in {\bf b}. Both conductance maps are taken at zero Zeeman energy.}
\vspace{-2mm}
\label{TS7}
\end{figure*}

\end{document}